\documentclass[useAMS,usenatbib,usegraphicx]{mn2e}

\defcitealias{Vivas2004}{V04}
\usepackage{times}
\usepackage[T1]{fontenc}
\usepackage{pslatex}
\usepackage{amsmath}
\usepackage{color,graphicx}

\newcommand{\lsim}{\raisebox{-.5ex}{$\,\stackrel{\textstyle <}{\sim}\,$}}
\newcommand{\gsim}{\raisebox{-.5ex}{$\,\stackrel{\textstyle >}{\sim}\,$}}
\footnotesize{\tiny}

\newcommand{\sqdeg} {${\rm deg}^2$ }
\newcommand{\rrab} {\mbox{RR\emph{ab}}}
\newcommand{\rrc} {\mbox{RR\emph{c}}}
\newcommand{\typeab} {\mbox{\emph{ab}}}
\newcommand{\typec} {\mbox{\emph{c}}}

\newcommand{\FeH} {[\mathrm{Fe}/\mathrm{H}]}
\newcommand{\VR} {\mathrm{\emph{V}}-\mathrm{\emph{R}}}
\newcommand{\VI} {\mathrm{\emph{V}}-\mathrm{\emph{I}}}
\newcommand{\RI} {\mathrm{\emph{R}}-\mathrm{\emph{I}}}
\newcommand{\BV} {\mathrm{\emph{B}}-\mathrm{\emph{V}}}
\title[The QUEST RR Lyrae Survey at Low Latitude]{The QUEST RR Lyrae Survey: \\ III. The Low Galactic Latitude Catalogue}
\author[C. Mateu et al.]{C.~Mateu,$^{1,2}$\thanks{cmateu@cida.ve} A.K.~Vivas,$^{1,3}$ J.J.~Downes,$^{1,2}$ C.~Brice\~no,$^{1,3}$ R.~Zinn,$^4$  G.~Cruz-Diaz $^5$ \\
$^{1}${{Centro de Investigaciones de Astronom\'{\i}a, AP 264, M\'erida 5101-A, Venezuela}}\\
$^{2}${{Escuela de F\'isica, Universidad Central de Venezuela, Apartado Postal 47586, Caracas 1041-A, Venezuela.}}\\
$^{3}${{Visiting Scholar at University of Michigan, Department of Physics, 500 East University, Ann Arbor, MI 48109, USA}}\\
$^{4}${{Department of Astronomy, Yale University, PO Box 208101, New Haven, CT 06520-8101, USA}}\\
$^{5}${{Centro de Astrobiolog\'ia, 28850 Torrej\'on de Ardoz, Madrid, Spain}}\\
}

\begin{document}

\maketitle

\label{firstpage}

\begin{abstract}
We present results for the QUEST RR Lyrae Survey at low galactic latitude, conducted entirely with observations obtained with the QUEST mosaic camera and the 1.0/1.5m J\"urgen Stock Schmidt telescope at the National Observatory of Venezuela. The survey spans an area of $476$ deg$^2$ on the sky, with multi-epoch observations in the $V$, $R$, and $I$ photometric bands for $6.5\times10^6$ stars in the galactic latitude range $-30\degr \leqslant b \leqslant +25\degr$, in a direction close to the Galactic Anticenter $190\degr \leqslant l \leqslant 230\degr$. The variability survey has a typical number of $30$ observations per object in $V$ and $I$  and $\sim25$ in $R$, with up to $\sim120-150$ epochs in $V$ and $I$ and up to $\sim100$ in $R$ in the best sampled regions. The completeness magnitudes of the survey are $V=R=18.5$ mag, and $I=18.0$ mag. We  identified 211 RR Lyrae stars, 160 \textit{bona fide} stars of type \typeab~and 51 candidates of type~\typec, ours being the first \emph{deep} RR Lyrae survey conducted at low galactic latitude.The completeness of the RR Lyrae survey was estimated in $\ga95$ per cent and $\sim85$ per cent for \rrab~and  \rrc~stars respectively. Photometric metallicities were computed based on the light curves and individual extinctions calculated from minimum light colours for each \rrab~star. Distances were obtained with typical errors $\sim7$ per cent. The RR Lyrae survey simultaneously spans a large range of heliocentric distances $0.5 \leqslant R_{hel}\mathrm{(kpc)} \leqslant 40$ and heights above the plane $-15\leqslant z\mathrm{(kpc)} \leqslant +20$, with well known completeness across the survey area, making it an ideal set for studying the structure of the Galactic thick disk.
\end{abstract}


\section{INTRODUCTION}\label{s:intro}
 
In the Milky Way (MW), the thick disk hosts a very old ($>10$ Gyr) and relatively metal-poor ($\FeH\sim-0.7$) stellar population \citep{Wyse2009,Reddy2008}, comprising $\sim10$ per cent or even up to $\sim20$ of the thin disk mass \citep[e.g.][]{Juric2008}. The Galactic thick disk therefore constitutes an important fossil record of the earliest stages of the formation of the Galactic disk, which could help understand the relevant mechanisms contributing to the formation of our Galaxy. From the perspective of galaxy formation, thick disks are remarkably relevant since these have proven to be an ubiquitous component of disk galaxies, as external galaxy surveys have shown that approximately $95$ per cent of disk galaxies contain thick disks \citep{Yoachim2006}. 

\begin{figure*}
\begin{center}
 \includegraphics[width=1.9\columnwidth]{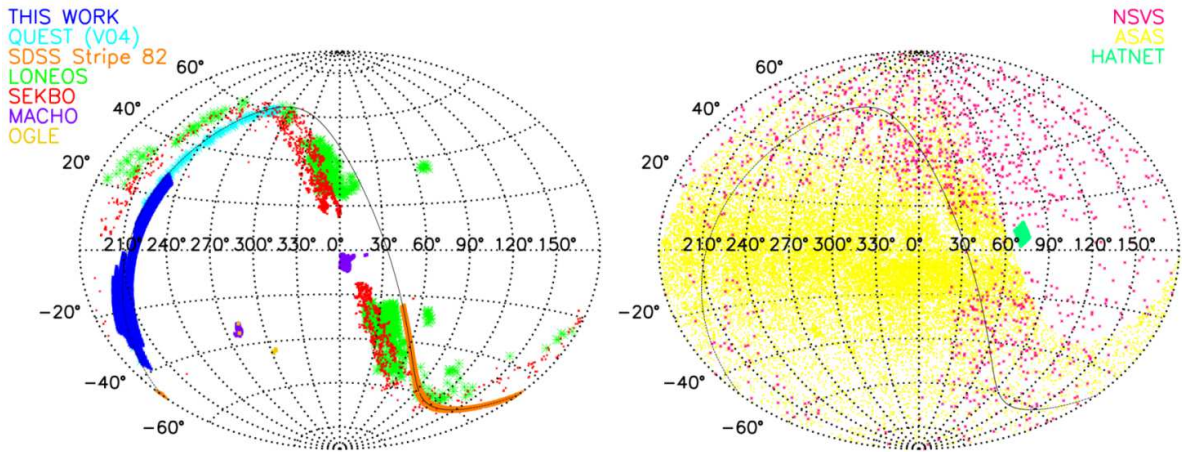}
 \caption{Aitoff projection map showing the footprints of recent large-scale RRLS surveys in Galactic coordinates. \emph{Left:} Deep surveys ($V>16$), \emph{(blue)} Present survey, \emph{(cyan)} QUEST halo RRLS survey, \emph{(orange)} SDSS Stripe 82, \emph{(green)} LONEOS,\emph{(red)} SEKBO, \emph{(purple)} MACHO, \emph{(dark yellow)} OGLE. \emph{Right:} Shallow surveys ($V<16$), \emph{(magenta)} NSVS, \emph{(yellow)} ASAS-3 and \emph{(light green)} HATNET. 
The solid black line corresponds to the Celestial Equator. Survey characteristics are summarized in Table \ref{t:survey_details}.}
\label{f:survey_coverage_aitoff}
\end{center}
\end{figure*}

On the other hand, despite many advances in the last few years, there is no consensus about the formation mechanism of the Galactic thick disk, or even on some fundamental properties of its structure, mainly due to observational difficulties. Scale length measurements for the MW thick disk range from $h_R\sim2$ to $4.7$ kpc \citep{Carollo2010,Larsen2003,CabreraLavers2005,Chiba2000}, which makes it unclear whether the thick disk is more radially extended than the thin disk or not, or how the density profile behaves at large radii. Another example are metal abundances, the iron abundance of thick disk stars seems to be fairly homogeneous in the vertical direction \citep[e.g.][]{Katz2011,Soubiran2003}, although \citet{Chen2011} recently report an appreciable metallicity gradient. Radial metallicity gradients have not been probed yet and there is also ongoing debate regarding the reach in $\FeH$ at both the low and high metallicity tails of the thick disk \citep{Bensby2007,Reddy2008,Reddy2005}. In a recent paper, \citet{Bovy2011,Bovy2011b} argue against the need of a thin/thick disk decomposition, proposing that the Galactic disk can be described as a series of simple stellar populations having scale lengths and scale heights which vary smoothly as a function of metallicity and $\alpha-$element abundance. This illustrates how the issue of Galactic thick disk structure is far from settled and reliable measurements on structural, kinematic and chemical properties are necessary to gain a better understanding of the MW's thick disk which in turn, will impose strong constraints on the possible contribution of the different formation mechanisms.

Observationally, disentangling the structure of the Galactic disk is a challenging task since the Sun is embedded in it. Numerous studies have been conducted at high galactic latitudes ($|b|\ga30\degr$) in order to study stars at a height above the Galactic plane $z\sim1-3$ kpc such that the thin disk density has decayed sufficiently, reducing contamination \citep[e.g.][]{Ojha2001,Siegel2002,Brown2008b,Wyse2009}, but also limiting the coverage in the radial direction. At intermediate to low latitudes ($|b|\la30\degr$) the problems of crowding and the high and variable extinction become important, and the contribution of the thin disk has to be modelled \citep{Robin1996,Larsen2003,Carollo2010,deJong2010}. 

RR Lyrae stars (RRLS) have several fundamental properties that offer advantages in dealing with these problems
inherent to low galactic latitude observations. RRLSs are \emph{luminosity and colour} standards. Since RRLSs are horizontal branch stars, they have a well known absolute magnitude which shows little spread \citep{Smith1995}, and allows the computation of distances with small uncertainties \citep[$\la10$ per cent,][hereafter \citetalias{Vivas2004}]{Vivas2004}. Aditionally, during the phase of minimum light of the pulsation cycle, RRLSs of type \typeab~have approximately the same effective temperature, and thus show very little dispersion in colour \citep{Sturch1966,Day2002}. This property enables the use of \rrab~stars as colour standards to measure extinctions up to the distance of each individual star; which is a crucial point in a low latitude survey since reddenning changes drastically along different lines of sight and has a strong dependence with distance. Also, since RRLSs trace old ($\ga$10 Gyr) and mainly metal-poor ($\FeH<-0.5$) stellar populations \citep{Smith1995,Demarque2000}, the expected contamination from the thin disk is negligible \citep{Martin1998}. Finally, RRLSs are pulsating stars with relatively short periods ($0.3-1.0$ d) and large light curve amplitudes ($V\sim0.3-1.2$ mag), which makes them easily identifiable by a photometric multi-epoch survey \citep[see e.g. \citetalias{Vivas2004},][]{Kinemuchi2006}.

\setcounter{table}{0}
\begin{table*}
 \begin{minipage}{160mm}
\caption{Characteristics of recent large-scale RRLS surveys}
\begin{footnotesize}
\begin{tabular}{lcccccl}
\hline
Survey & Filters & Area (\sqdeg) & Completeness & Telescope & Observatory & Reference \\
\hline
QUEST  & $VRI$ & $476$ & $V<19$\,\,\,\, & $1.0$ m Schmidt & NOV & This Work \\
QUEST  (V04) & $V$ & $380$ & $V<19$\,\,\,\, & $1.0$ m Schmidt & NOV & \citetalias{Vivas2004} \\
NSVS         & ROTSE-NT & $\sim31000$       & $V<14$\,\,\,\, & $4\times200$ mm ROTSE-I  & Los Alamos & \citet{Kinemuchi2006} \\
ASAS-3      & $V$     & $\sim31000$ & $V<14$\,\,\,\, & $2\times200$ mm ASAS & Las Campanas & \citet{Pojmanski2002} \\
LONEOS    & LONEOS-NT & $1430$        & $V<18$\,\,\,\, & $0.6$ m Schmidt & Lowell  & \citet{Miceli2008} \\
SEKBO      & $B_MR_M$  & $1675$    & $V<19.5$ & $1.27$ m MACHO & Mount Stromlo & \citet{Keller2008} \\ 
HATNET    & $I$ & $67$ & $V<15$\,\,\,\, & $11$\,\,\,\, cm HATNET & Fred L. Whipple & \citet{Hartman2004} \\
OGLE        & $I$ & $0.87$ & $V<20.5$ & $1.3$ m Warsaw & Las Campanas & \citet{Udalski1998} \\
SDSS Str82 & $ugriz$  & $249$     & $V<22$\,\,\,\, & $2.5$ m SDSS & Apache Point & \citet{Watkins2009} \\
 & & & & & & \citet{Sesar2010}\\
MACHO     & $B_MR_M$  & $\sim75$ & $V<20$\,\,\,\, & $1.27$ m MACHO & Mount Stromlo & \citet{Kunder2008}, \\
 & & & & & & \citet{Alcock2003}\\
\hline
\end{tabular}
\end{footnotesize}
\label{t:survey_details}
\end{minipage}
\end{table*}

RRLSs have been extensively used as tracers of the Galactic halo by numerous surveys which have studied their spatial distribution in a wide range of distances, from very near the Galactic centre up to large distances $R_{gal}\ga100$ kpc. This is illustrated in Figure \ref{f:survey_coverage_aitoff} which shows, in an Aitoff projection map in galactic coordinates, the footprint of recent large-scale RRLS surveys including the present one. Table \ref{t:survey_details} summarizes the characteristics of each survey. 

Unlike in the Galactic halo, the distribution of RRLSs in the thick disk has been less thouroughly studied, in particular due to the lack of deep RRLS surveys at low/intermediate galactic latitudes. As illustrated in Figure \ref{f:survey_coverage_aitoff}, the Galactic Plane area $|b|<20\degr$ has only been covered by the large-scale yet shallow ($V<14$ or $R_{hel}\la 5$ kpc) ASAS-3 \citep{Pojmanski2002} and NSVS \citep{Kinemuchi2006} surveys, as well as the compilations of nearby RRLS ($V\lsim13$) from \citet{Layden1994,Layden1995} and \citet{Maintz2005}; having been avoided by deep surveys ($16<V<20$). This makes ours, the first deep large-scale RRLS survey conducted at low Galactic latitudes. Furthermore, our survey probes the outer regions of the thick disk, outside the solar circle. In particular, \citet{Layden1995,Maintz2005b} and \citet[][NSVS]{Kinemuchi2006} have studied the distribution of thick disk RRLSs in the vertical direction, but the limited coverage at low galactic latitude prevented the determination of the scale length as well as the exploration of deviations from an exponential profile such as the \emph{warp} and \emph{flare}, which have been observed in the Galactic thin and thick disks with other tracers \citep[e.g.][]{LopezCorredoira2002,Momany2006,Hammersley2011}. For these reasons it was essential to conduct a deep RRLS survey at low galactic latitude, which provided an ample spatial coverage both in the radial and vertical directions.

In this paper we will present the catalogue of the QUEST\footnote{Acronym for the `QUasar Equatorial Survey Team' } RR Lyrae survey of the thick disk, which spans a total area of $476$ \sqdeg at $-30\degr<b<+25\degr$, with multi-epoch $V,R,I$ observations obtained with the $1.0/1.5$m J\"urgen Stock  Schmidt telescope and the QUEST mosaic camera, at the National Astronomical Observatory of Venezuela (NOV). In Section \ref{s:survey} we describe the observations and spatial and time coverage of the survey, as well as the photometric processing, error estimation and calibration. In Section \ref{s:rrl_search} we describe the procedures used in the identification of RRLSs, we estimate the completeness and possible contamination of the survey and we describe the computation of periods, amplitudes, reddenings, photometric metallicities and distance for the identified RRLS. In Section \ref{s:conclusions} we summarize the characteristics of our survey. The computation and analysis of RRLS density profiles for the Galactic thick disk and halo will be presented in an upcoming paper of the series.

\section{THE SURVEY}\label{s:survey}

\subsection{The Data}

The present survey spans a total area of $476$~deg$^2$ and makes use of archive observations obtained between late 1998 and mid-2008 with the QUEST camera and the $1.0/1.5$ m J\"urgen Stock Schmidt telescope at the NOV in Llano del Hato, Venezuela. The camera is a 16 CCD mosaic, with $2048 \times 2048$ pixel CCDs layed out in a $4 \times 4$ array, having a field of view of $2\fdg3 \times 2\fdg3$ and a scale of $1\farcs02$ pixel$^{-1}$. The QUEST camera was designed to operate optimally in \emph{drift-scanning} mode  and can be fitted with up to four different filters at a time. For a more detailed description of the camera and the drift-scanning technique see \citet{Baltay2002} and \citetalias{Vivas2004}.

For the present survey we used archive drift-scan observations at low Galactic latitudes ($|b|\la30\degr$) approximately in the direction towards the Galactic anticentre. These observations were obtained by different observational projects with very different goals, but since the equipment and method of observations are the same as the original QUEST RR Lyrae survey \citepalias{Vivas2004}, we have decided to keep that name for the present survey.  Many of the observations come indeed from the QUEST (quasar) survey (Baltay et al. 2002; Rengstorf et al. 2004, V04), and there is also an important number of scans from the CIDA\footnote{Acronym for the `Centro de Investigaciones de Astronom\'ia'} Equatorial Variability Survey \citep{Briceno2003}. The latest spans the equatorial region in the full range of right ascension and has thoroughly surveyed the Orion star forming region, resulting in a large number of observations in this area  \citep{Briceno2005,Downes2012}. Our sample is restricted to observations in the range $60\degr \leq\alpha\leq135\degr$ in right ascension (corresponding to the range $190\degr \leq l \leq 230\degr$ in Galactic longitude), where the Galactic Plane crosses the celestial equator and drift-scan observations in the declination range $|\delta|\leq6.5\degr$ are available. 

The observations were obtained using different filter combinations and have different spatial coverage and cadence, all of which changed over time, resulting in a highly inhomogeneous sample. The survey is composed by 452 drift-scans corresponding to $19,238$ hours of observation and $6.3$ Tb of data in the $V$, $R$ and $I$ photometric bands, obtained in a total of $427$ nights between December 1998 and June 2008. Although ideally an RRLS survey would use a blue filter, since the amplitudes of variation of RRLS would be larger, we did not include these observations in our sample since these were very few due to the low sensitivity of QUEST CCDs in this band.
 
The present survey includes the low galactic latitude observations from the RRLS survey of \citetalias{Vivas2004}, which consisted of QUEST drift-scans at $\delta=-1\degr$ obtained between December 1998 and April 2001. This results in an overlap area of $90$~\sqdeg at both ends of the present survey (see Figure \ref{f:survey_coverage_aitoff}) in the right ascension ranges $60\degr \leq\alpha\leq90\degr$ and $120\degr \leq\alpha\leq135\degr$ and the declination range $-2\degr \leq\delta\leq0\degr$. New observations (post-2001) were added in this area and the existing ones from \citetalias{Vivas2004} were reprocessed using the procedures described in the following section, which were devised to make an optimal use of data in the three available filters $V,R,I$. Finally, data in the range $60\degr\leq\alpha\leq90\degr$ are also part of the CIDA Deep Survey of Orion \citep[in preparation]{Downes2012}. 

\subsection{Data Reduction and Photometry}

In the following sections we will describe the procedures used in the object detection, astrometric reduction, aperture and PSF photometry.

\subsubsection{Object Detection, Astrometry and Aperture Photometry}

Each individual drift-scan observation was reduced using \emph{Offline}, the standard pipeline developed by the QUEST colaboration \citep{Baltay2002}. The \emph{Offline} pipeline automatically performs basic reductions, object detection, aperture photometry and astrometric reductions for each image frame in a drift-scan observation. This processing was conducted independently for the images obtained by each of the camera's CCDs , resulting in 16 independent catalogues having position in celestial equatorial coordinates, photometric magnitude and error for each point-source, as well as photometric parameters such as FWHM, ellipticity, sky background flux and error, etc.

Image reduction consists in standard bias, dark and flat-fielding corrections. Object detection is made at the 3-$\sigma$ level and aperture photometry is performed over all detected point-sources with an aperture radius of one FWHM. The astrometric matrices are computed using the USNO-A2 \citep{Monet1998} astrometric catalogue as a reference. The astrometric solutions computed by \emph{Offline} have a typical accuracy of $\sim0\farcs15$, which is sufficient for the cross-identification of an object in the different catalogues.

\subsubsection{PSF Photometry}

\begin{figure}
\begin{center}
 \includegraphics[width=\columnwidth]{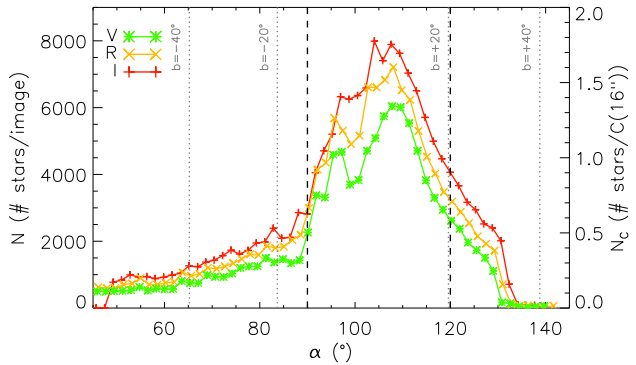}
\caption{Number of stars per image as a function of $\alpha$, for a drift-scan centered at $\delta=-3\degr$, with  filters \emph{V}, \emph{R} e \emph{I} ($\ast,\times,+$ respectively). The right-hand vertical axis indicates the typical number of stars $N_C$ inside a circle of radius $16\arcsec$ ($C(16\arcsec)$). Dotted grey lines indicate the regions corresponding to galactic latitudes $b=\pm20\degr,\pm40\degr$. The black dashed lines delimit the region were PSF photometry was computed.}
\label{f:ndetecc_vs_ra}
\end{center}
\end{figure}

Since our drift-scan observations include very crowded regions at low Galactic latitude, the use of PSF photometry was necessary. Figure \ref{f:ndetecc_vs_ra} shows a plot of the typical number of stars per image in the $V,R$ and $I$ bands, as a function of right ascension $\alpha$ for a drift-scan at $\delta=-3\degr$ which spans the entire right ascension range covered by this survey. The right-hand vertical axis shows the scale in terms of $N_C$ the number of stars in a circle with a $16\arcsec$ radius (which is twice the aperture diameter in an observation with a bad seeing of $4\arcsec$). The plot illustrates how the number of objects per image increases when approaching the Galactic Plane. In particular $N_C>1$ in the range $95\degr \la \alpha \la 115\degr$, meaning that the typical separation between two objects in an image is less than twice the aperture diameter and, therefore, aperture photometry for most stars will be contaminated by neighbours. 
Therefore, the computation of PSF photometry was restricted to the right ascension range $90\degr \leqslant \alpha \leqslant 120\degr$, in which $N_C\ga0.8$ (indicated in Figure \ref{f:ndetecc_vs_ra} with dashed black lines). In low density areas, consistency was checked by computing both aperture and PSF photometry.

PSF photometry was obtained using the standard \emph{DAOPHOT} \citep{Stetson1987} routines implemented in IRAF\footnote{IRAF is distributed by the National Optical Astronomy Observatories, which are operated by the Asociation of Universities for Research in Astronomy Inc., under cooperative agreement with the National Science Foundation}, tied together in an automated pipeline called \emph{OfflinePSF}, which is customized for QUEST drift-scan observations. 

\emph{OfflinePSF} includes as a first step, prior to the computation of PSF photometry, the recentering and recomputation of aperture photometry for all point-sources previously detected with \emph{Offline}. This is necessary since initial Offline detections have centering errors of $\sim 1$ pixel, which are reduced to $\sim 0.1$ pixel by using PHOT's recentering options. Using the recalculated centroids and aperture photometry magnitudes, PSF photometry is obtained according to an iterative scheme where, for each individual image, input parameters such as the fit radius $r_{fit} = 1\times FWHM$ and the model PSF radius $r_{psf} = 3 \times r_{fit}$ are computed independently. 
Then, from 90 to 120 PSF stars are selected, uniformly distributed accross the image, without neighbours within a radius $r_{psf}$ and, for distances less than $2r_{psf}$ requiring neighbourghs be at least one magnitude fainter and there are no bad pixel columns near. An initial PSF model is made with the DAOPHOT PSF task, with a second order spatial variation across the image. The PSF model is fit around PSF stars with the NSTAR and SUBSTAR tasks, PSF stars with newly detected neighbourghs are discarded and the PSF model recomputed. The PSF model is then fit to all point-sources in the entire image using ALLSTAR and new objects are detected if present in the resulting subtracted frame, iterating twice more over these two steps. 

The \emph{Offline} and \emph{OfflinePDF} processes result, for each drift-scan, in a photometric catalogue per CCD, containing the equatorial coordinates $\alpha,\delta$, instrumental aperture and PSF magnitudes and errors, for each point-source object detected in the full drift-scan. 

\subsubsection{Photometric Calibration}

The photometric calibration was done in two stages: the calibration of a \emph{master catalogue} by means of second order photometric standards and the normalization of individual drift-scan observations to this master photometric catalogue.

\begin{figure}
\begin{center}
\includegraphics[width=\columnwidth]{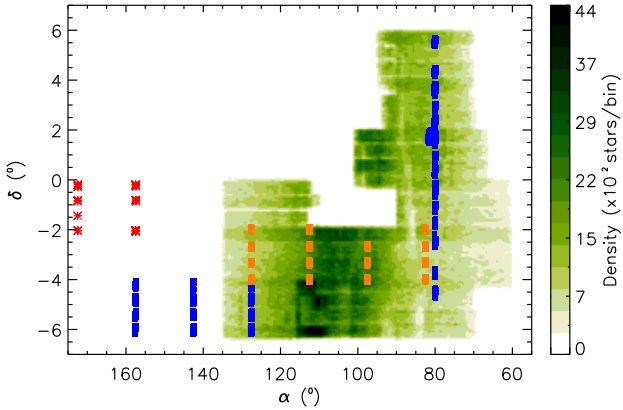}
 \caption{Density map of stars in the master photometric catalogue and spatial distribution of second order photometric stantards used for photometric calibration. The gray scale represents stellar density as indicated by the colour bar, per $0.24$ deg$^2$ bin (equivalent to half the area of a QUEST CCD). The dots indicate the positions of second order photometric standards established by different authors: (blue) \citet{Downes2012}, (red) \citetalias{Vivas2004} and (orange) this work.}
\label{f:secstand_map}
\end{center}
\end{figure}

The \emph{master catalogue} is composed of 3 to 4 individual scans per declination stripe, observed with good and constant seeing and sky background throughout the whole duration of the scan.
The combination of these scans allowed for a much more homogeneous master catalogue, than would have been obtained if only one scan had been used, due to empty regions caused by bad columns and defective regions of the CCDs. The \emph{master catalogue} obtained contains 6,513,705 objects and its coverage is illustrated in the density map shown in Figure \ref{f:secstand_map}. 

The instrumental magnitudes for each object in the \emph{master catalogue} were calibrated using second order photometric standards. A total of 4526 photometric standards were used, spanning the magnitude range $10<V<17$ with colours $-0.2<\VR<1.2$ and identified as non-variable stars in any one of the $VRI$ bands, with a spatial distribution illustrated in Figure~\ref{f:secstand_map}. Photometric standard stars indicated by black dots (2621 in total) were set up by \citet{Downes2012} using observations obtained with the 1.2 m telescope and the Keplercam camera, at the Fred L. Whipple Observatory, Mt. Hopkins, US. The asterisks indicate the 28 standards having $\delta=-1\degr$ and $\alpha<175\degr$, established by \citetalias{Vivas2004} with observations from the 1.0m YALO\footnote{Acronym for 'Yale Aura Lisbon Ohio'} telescope at the Cerro Tololo Interamerican Observatory (CTIO), Chile. Finally, as part of this work, we established 1877 second order standards in the range $-4\degr\leq\delta\leq-2\degr$ (denoted in Figure \ref{f:secstand_map} with dark gray dots), using observations obtained during December 2004, with the 0.9 m SMARTS\footnote{Acronym for 'Small and Moderate Aperture Reasearh Telescope System'} telescope at CTIO. Data for these 1877 second order standards is summarized in Table \ref{t:secstands}. The photometric calibration was performed independently in declination stripes having the width of a QUEST CCD  and in 1-hour-long intervals in right ascension where the distribution of standards allowed it. The catalogue regions centered around $\delta=-1\degr$ with $110\degr\leq\alpha\leq135\degr$ and $\delta<-5\degr$ with $\alpha>130\degr$  lacked photometric standard stars, so these were calibrated using long reference scans which extended as far as the area covered by the \citetalias{Vivas2004} standards (asterisks) and \citeauthor{Downes2012} (black dots) respectively.

\begin{table*}
 \begin{minipage}{120mm}
\caption{Magnitudes for second order photometric standards.}
\begin{footnotesize}
\begin{tabular}{cccccc}
\hline
ID & $\alpha$ &  $\delta$  & $V$ & $R$ & $I$ \\
    & $(\degr)$& $(\degr)$ & (mag) & (mag) & (mag) \\
\hline
ss4000 &   82.401749  &   -2.04192 & 13.446 $\pm$  0.003 & 12.238 $\pm$  0.003 & 12.822 $\pm$  0.003 \\
ss4001 &   82.420319  &   -2.10225 & 14.386 $\pm$  0.005 & 13.201 $\pm$  0.005 & 13.800 $\pm$  0.004 \\
ss4002 &   82.431961  &   -2.04361 & 16.905 $\pm$  0.020 & 14.526 $\pm$  0.009 & 15.849 $\pm$  0.012 \\
ss4003 &   82.436493  &   -2.11058 & 16.630 $\pm$  0.018 & 15.197 $\pm$  0.014 & 15.912 $\pm$  0.012 \\
ss4004 &   82.437607  &   -2.03612 & 16.784 $\pm$  0.019 & 15.329 $\pm$  0.015 & 16.046 $\pm$  0.013 \\
\hline
\end{tabular}
\end{footnotesize}
\label{t:secstands}
\footnotetext{Note: Table \ref{t:secstands} is published in its entirety in the electronic edition of the journal. A portion is shown here for guidance regarding its form and content.}
\end{minipage}
\end{table*}

The photometric solutions for each scan were obtained using zero-point and $\VR$ or $\VI$ colour terms, depending on the filters available for each observation. A mean photometric precision of $\sim0.021$ mag was achieved for the $V,R$ and $I$ bands.

 \subsubsection{Normalization}\label{s:norm}

 The master photometric catalogue, described in the previous subsection, provides the calibrated reference photometry for all survey objects. The individual observations, used to construct time series for each object (see Sec. \ref{s:vars}) used in the RRLS search, were calibrated by normalizing each scan to the master reference catalogue.

The normalization process, taken from \citetalias{Vivas2004}, consists in identifying all objects of a particular scan in the reference master catalogue and computing the mean magnitude residuals in both, throughout the whole scan in  $0\farcm25$ bins in right ascension. Since most stars ($\ga95$ per cent) are non-variable and the number of stars in each bin is sufficiently large ($\sim1500-3000$), the average residuals obtained result in a precise estimation of the zero point needed to calibrate each bin of the scan. This normalization process allows for the correction of systematic trends in the magnitude residuals caused by changes in the atmospheric conditions during each observation, which in our case is critical given that a single drift-scan can be up to 7 to 8 hours long \citepalias[see Figure 3 in][for an example of the normalization process]{Vivas2004}. The typical resulting standard deviation of the residuals in each bin was $\sim0.015$ mag, giving zero point errors $\sim0.002$ mag.

\subsubsection{Photometric Errors}\label{s:photerrs}

\begin{figure}
\begin{center}
\includegraphics[width=\columnwidth]{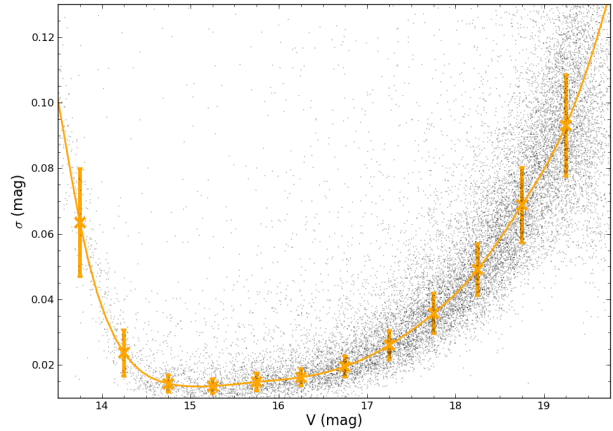}
 \caption{Magnitude standard deviation $\sigma$ as a function of mean V magnitude, for stars in a typical $0.5\degr \times 15\degr$ region centered around $(\alpha,\delta)=(85\degr,+3\degr)$, with a mean number of $\sim25$ observations per star. The solid line indicates the polynomial fit of the average $\sigma$ in 0.5 mag bins, as a function of mean magnitude. Error bars indicate the standard deviation around each of these averages.}
\label{f:err_vs_mag}
\end{center}
\end{figure}

Taking advantage of the fact that each star in our survey has multiple observations in each filter, it is suitable to use the standard deviation $\sigma$ of magnitudes of non-variable stars as the representative photometric error in the given filter. Since the majority of stars in any given field are non-variable, these give rise to the main trend in a plot of $\sigma$ versus mean magnitude, as seen in Figure~\ref{f:err_vs_mag}. 

In the figure the crosses show the average standard deviation computed as a function of mean magnitude in $0.5$ mag bins using an iterative $3\sigma-$clipping. The sharp increase in $\sigma$ for objects brighter than $V\sim14.5$ mag is due to objects saturated in most observations. This behaviour provided us with a means to empirically determine the saturation magnitude and, for each object, remove the individual observations brighter than this limit. The solid line indicates a polynomial fit of the average $\sigma$ versus magnitude (crosses) and the error bars  correspond to the standard deviation around this average. The polynomial fit consequently provides the typical magnitude standard deviation due to photometric uncertainties only, as a function of an object's mean magnitude, which we assumed as its photometric error.

For each survey star we computed the corresponding photometric error as the the typical magnitude standard deviation given by the polynomial fit, for the corresponding mean magnitude.
This error analysis was conducted independently for observations in each of the $V$,$R$ and $I$ filters and dividing the entire survey in separate regions of $\Delta\delta=0.5\degr$ times $\Delta\alpha=15\degr$.
 
\subsection{The Photometric Catalogue}\label{s:photcat}

\begin{figure*}
\begin{center}
\includegraphics[width=2.2\columnwidth]{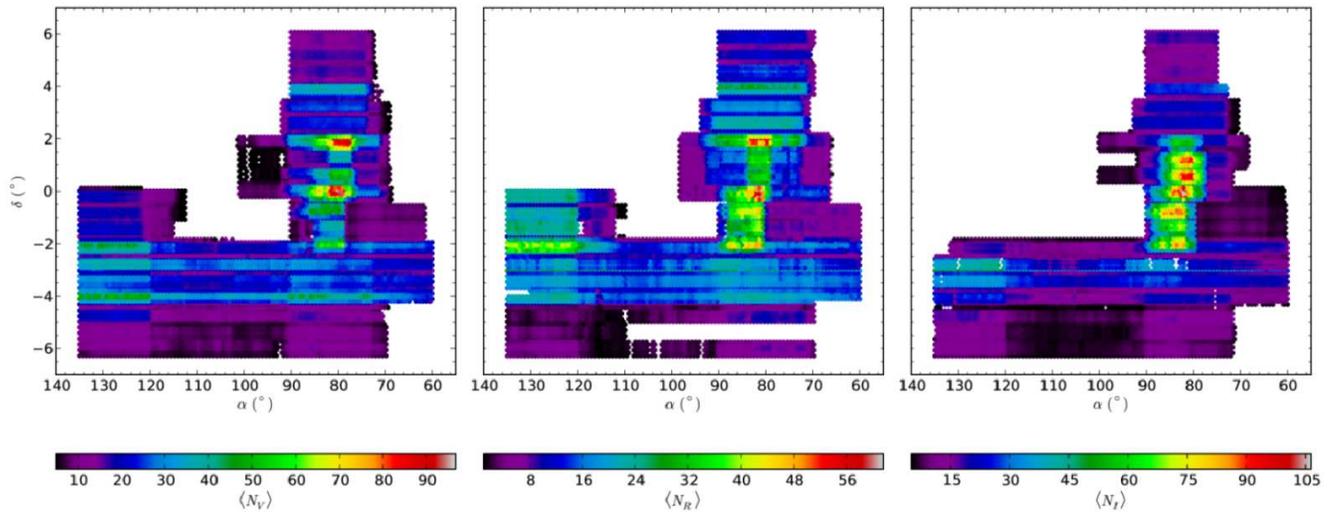}
 \caption{Spatial coverage of the survey in the \emph{V} (\emph{top}), \emph{R} (\emph{middle}) e \emph{I} (\emph{bottom}) bands. The colourbars indicate the mean number of epochs per object in each filter.}
\label{f:map_nvri}
\end{center}
\end{figure*}

The final photometric catalogue spans a total area of $476$ sq. deg. having $6,513,705$ objects with multi-epoch observations in the $V$, $R$, and $I$ bands. Table \ref{t:survey_mags} summarizes the saturation ($m_s$), completeness ($m_c$) and limiting ($m_l$) magnitudes in the three photometric bands, as well as the typical errors corresponding to these magnitudes. 

The spatial coverage of the survey as well as the typical number of epochs per object, in each of the photometric bands, are illustrated in Figure \ref{f:map_nvri} in equatorial coordinates. As can be seen in the figure, the number of observations per object goes from $\sim10$ up to $\sim150$ in the $V$ and $I$ bands and up to $\sim60$ in the R band. The number of epochs per star per filter is quite inhomogeneous throughout the survey, as illustrated by Figure \ref{f:map_nvri}, nevertheless most objects have at least $10$ observations in two photometric bands; while objects in the thoroughly sampled Orion region have more than $\sim30$ observations in each of the $V$, $R$ and $I$ bands.

\begin{table}
 \begin{minipage}{80mm}
\caption{Saturation, completeness and limiting magnitudes of the survey.}
\begin{footnotesize}
\begin{tabular}{ccccccc}
\hline
Filter & $m_{s}$ & $\Delta m_{s}$ & $m_{c}$ & $\Delta m_{c}$ & $m_{l}$ & $\Delta m_{l}$ \\
         & (mag) & (mag) & (mag) & (mag) & (mag) & (mag)  \\
\hline
\emph{V}  & $14.0$ & $0.018$ & $18.5$ & $0.064$ & $19.70$ & $0.15$ \\
\emph{R}  & $14.0$ & $0.018$ & $18.5$ & $0.072$ & $19.75$ & $0.15$ \\
\emph{I}   & $13.5$ & $0.020$ & $18.0$ & $0.078$ & $18.80$ & $0.15$ \\
\hline
\end{tabular}
\end{footnotesize}
\label{t:survey_mags}
\end{minipage}
\end{table}

\section{THE RR LYRAE SEARCH}\label{s:rrl_search}

\subsection{Light Curve Simulations}\label{s:synth_rrls}

\begin{table*}
 \begin{minipage}{120mm}
\caption{Parameters used in the simulation of \rrab\, and \rrc\, light curves.}
\begin{footnotesize}
\begin{tabular}{lccc}
\hline
Parameter & \rrab  & \rrc  & Reference\\
\hline
$\langle P\rangle$    & $0.539$ d & $0.335$ d & \citetalias{Vivas2004} \\ 
$\sigma P$               & $0.09$ d   & $0.07$ d  & \citetalias{Vivas2004}\\
\hline
$\langle {\rm Amp}V\rangle$  & $1.04$ mag  & $0.536$ mag & \citetalias{Vivas2004}\\
$\sigma_{\mathrm{Amp}V}$         & $0.24$ mag  & $0.13$ mag  & \citetalias{Vivas2004}\\
$\mathrm{Amp}R$ & \multicolumn{2}{c}{$0.74064\times\mathrm{Amp}V\,+ 0.0361$} & \citet{Alcock2003} \\
$\mathrm{Amp}I$ & \multicolumn{2}{c}{$0.66800\times\mathrm{Amp}V\,+ 0.0501$} & \citet{Dorfi1999} \\
\hline
$\langle V \rangle$   & \multicolumn{2}{c}{$13.0-20.0$ } & $\cdots$ \\
$(\langle \VR \rangle,\sigma_{\VR})$ & \multicolumn{2}{c}{$(0.30;0.26)$} &  \citet{Alcock2003} \\
$(\langle \VI \rangle,\sigma_{\VI})$ & \multicolumn{2}{c}{$(0.42;0.20)$} &  \citet{Alcock2003} \\
\hline
\end{tabular}
\end{footnotesize}
\label{t:RRsim_params}
\end{minipage}
\end{table*}

The present survey has very inhomogenous time sampling and number of epochs in the different
filters, which in turn change in different regions of the survey. These differences can have a 
sizeable impact in the completeness of the resulting RRLS sample, which must be properly characterized. 
We used light curve simulations as a tool to test and fine-tune the variable star detection and period searching techniques used to identify RRLS stars (Sec. \ref{s:vars}~and~\ref{s:period_search}), as well as to characterize the completeness of the \rrab~and \rrc~ samples as a function of position in the survey (Sec. \ref{s:completeness}).

A total of $425,000$ synthetic light curves were generated for RRLS of each type, \typeab~ and \typec, and the same number for non-variable stars. The time sampling and photometric errors  as a function of magnitude (see Sec. \ref{s:photerrs}) were taken from actual objects uniformly distributed throughtout the entire survey area. 
This ensures a truly representative realization of the time spacing, the number of observations per filter and the number of quasi-simultaneous observations in different filters, for each synthetic object. This point is of particular importance given the differences both in the cadence and spatial coverage in the different photometric bands as a function of position, as illustrated in Figure \ref{f:map_nvri}.

The parameters used to create synthetic light curves for \rrab~and \rrc~stars are summarized in Table \ref{t:RRsim_params}. We generated synthetic stars with random mean $V$ magnitudes $\langle V \rangle$ uniformly distributed in the range $[13.0,20.0]$. For each object, the corresponding mean magnitudes $\langle R\rangle$ and $\langle I \rangle$ were computed from the mean $V-R$ and $V-I$ colours, randomly drawn from gaussian distributions with means and standard deviations taken from \citet{Alcock2003}, indicated in Table \ref{t:RRsim_params}. The pulsation period $P$ of each object was randomly drawn from a gaussian distribution and the $V$-band amplitude computed from the (Oosterhoff I) Period-Amplitude relation from \citet{Cacciari2005}, with random gaussian noise with sigma $0.12$ mag, for \rrab~stars. For \rrc~stars, since the Period-Amplitude relation is weak, the V-band amplitude was randomly drawn from a gaussian distribution assuming a mean a standard deviation taken from \citetalias[][see Table \ref{t:RRsim_params}]{Vivas2004}. This way our simulated population of RRLS shows the period-amplitude trend for Oosterhoff I \rrab~stars \citep[which represent $\sim80\%$ of all \rrab~stars in the Galactic halo,  e.g.][]{Sesar2010}, albeit not the Oosterhoff dicotomy. Note that although we do not simulate the Oosterhoof dicotomy, nor the Blazhko effect \citep[an amplitude modulation known to occurr in some RRLS, e.g.][]{Kovacs2009}, we do not expect they will affect our completeness estimates for the survey. For a given period, Oosterhoff II stars would have even larger amplitudes than Oosterhoff I ones making them even easier to identify. On the other hand, the amplitude of RRLS is large enough such that the (relatively small) variations due to Blazhko effect will only translate in the light curves appearing slightly more noisy. 

 Given the V-band amplitude (Amp$V$), the corresponding amplitudes in the $R$ and $I$ bands were computed according to the linear relationships from \citet{Dorfi1999} and \citet{Alcock1999,Alcock2003} respectively, including the gaussian uncertainty estimated for these relationships. This procedure correctly reproduces the amplitude ratios observed for RRLS in different photometric bands \citep[e.g.][]{Smith1995,Sterken2005}. Finally, an initial phase offset randomly drawn from a uniform distribution in the $\phi\in[0,1)$ range, was added to the synthetic light curve of each object which in turn was sampled from \rrab~or \rrc~light curve templates from \citet{Layden1998} and \citet{Vivas2008} respectively, in order to obtain the magnitudes corresponding to each epoch of observation (in each filter). A random gaussian photometric uncertainty was added to each of the synthetically produced magnitudes, using the error vs magnitude curves for each filter derived for the survey (see Sec. \ref{s:photerrs}).

Time series data were also generated for synthetic non-variable stars with $\langle V \rangle \in [13.0,20.0]$. Mean $\langle R\rangle$ and $\langle I \rangle$ magnitudes were computed from $V-R$ and $V-I$ colours, randomly drawn from uniform distributions in the ranges $[-0.1,1.5]$ and $[-0.2,3.0]$ respectively, which correspond to the entire $V-R$ and $V-I$ colour range for main sequence stars according to \citet{Kenyon1995}.

\subsection{Variable star identification}\label{s:vars}

\begin{figure*}
\begin{center}
 \includegraphics[width=1.\columnwidth]{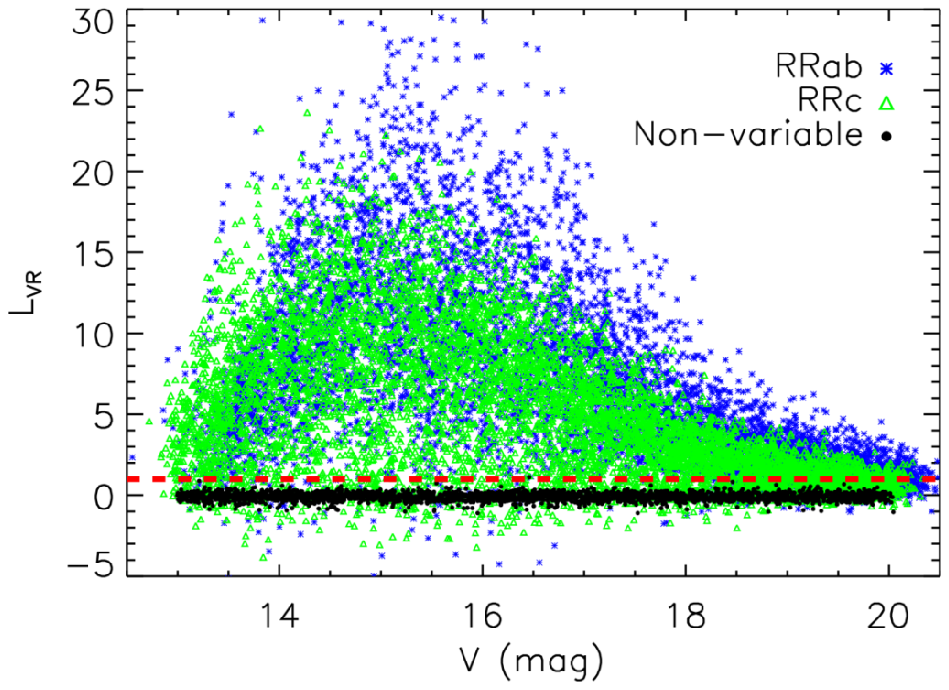} 
 \includegraphics[width=1.\columnwidth]{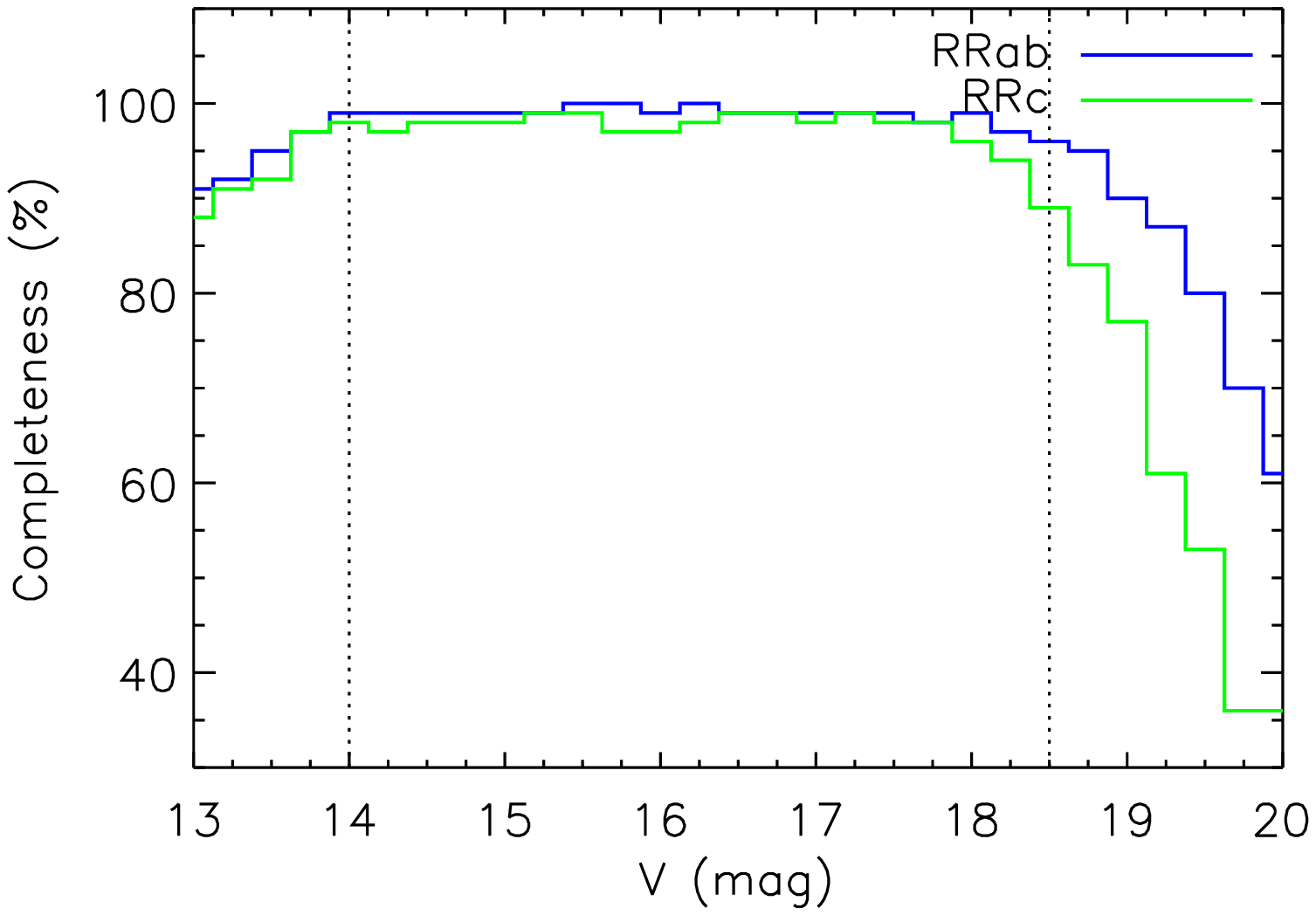} 
 \caption{\emph{Left:} $L_{V\!R}$ versus \emph{V} magnitude, for synthetic non-variable \emph{(black)}, \rrab~\emph{($\ast$)} and \rrc~\emph{($triangle$)} stars. The dashed line indicates the threshold chosen for the selection of variable stars. \emph{Right:} Completeness of the variable star identification, estimated from synthetic light curves of \rrab~\emph{(blue)} and \rrc\, \emph{(green)} stars. The dotted lines indicate the survey's saturation and completeness magnitudes in the V-band (see Table \ref{t:survey_mags})}
 \label{f:L_completeness}
\end{center}
\end{figure*}

The first step in the search for RRLS consisted in identifying variable stars, by means of the variability indices defined by \citet{Stetson1996}. 

The $J$ and $K$ variability indices were initially defined by \citet{Welch1993} and later modified and combined by \citet{Stetson1996} in the definition of the $L$ index, specifically designed for the automated identification of pulsating variables and the reduction of contamination due to spurious variables caused by deviant photometric observations. The definition of the $L$ index uses the fact that, for pulsating stars, quasi-simultaneous observations in two photometric bands must experience deviations (with respect to the corresponding means) which show a positive correlation, as opposed to non-variable stars for which this deviations would be uncorrelated. 

The QUEST drift-scans comprising this survey have data in up to four different filters (see Sec. \ref{s:photcat}) obtained almost simultaneously, since the mean separation between observations in two different filters is $\Delta t\sim2-6$ min (given by the time it takes an object to cross the different CCD rows of the camera), which compared to the typical period of an RRLS star ($P\sim0.55$ d) can be considered negligible ($\Delta t/P\sim7\times10^{-3}$). This makes it an ideal set of observations to use \citeauthor{Stetson1996}'s variability indices.

Candidate variable stars were identified based on the three \citeauthor{Stetson1996} $L$ indices  that can be computed with our observations, i.e. $L_{VR},L_{VI}$ and $L_{RI}$. Given that the survey coverage and sampling are very inhomogeneous in the different filters (see Figure \ref{f:map_nvri}), the criteria used in identifying variable stars needed to be sufficiently flexibe in order to consider e.g. objects which have observations in any two bands and not the third, yet without seriously reducing the completeness in the process. Therefore, we selected as candidate variable stars, those objects having at least one of the three possible $L$ indices greater than a threshold value of $1$, slightly higher than the threshold value $0.9$ used by \citet{Stetson1996} in the identification of Cepheid variables. This threshold value was chosen based on results from light curve simulations of \rrab, \rrc~and non-variable stars with the survey's time sampling.

A plot of $L_{VR}$ as a function of $V$ is shown in Figure \ref{f:L_completeness} (left). As expected, the larger amplitude \rrab~stars (blue dots) have on average large $L$ values than the lower amplitude \rrc~stars (green dots), while non-variable (black dots) stars have $L$ values around $0$. The criteria used to select variable star candidates (at least one $L$ index higher than 1, see Sec. \ref{s:vars}) ensures there will be negligible contamination from non-variable and low-amplitude ($\la0.2$ mag) variable stars, while still recovering most variable stars, regardless of whether these have a small number of observations ($\la10-12$) in some filters. The completeness of the variable candidate sample achieved using this criteria is illustrated as a function of mean $V$ magnitude in Fig \ref{f:L_completeness} (right). Within the saturation and completeness magnitude limits (dashed lines), the completenetess in the variable star identification was estimated in $99.1$ per cent for \rrab~and $98.3$ per cent for \rrc~stars.

Using this criteria we obtained a sample of $486,766$ candidate variable stars, corresponding to $7$ per cent of the total number of stars in the survey. 

\subsection{RR Lyrae identification}\label{s:rrl_ident}

\subsubsection{Period Search}\label{s:period_search}

The search for RRLS was conducted using a version of the \citet*{Lafler1965} method for the identification of periods of variable stars, modified by \citet{Stetson1996} and extended in the present work in order to incorporate the information of multi-filter time series data. 

The method consists in computing, for a series of trial periods, the parameter

\begin{equation}\label{e:s_lkstetson}
S(M)=\frac{\sum_i^{N_M} \omega(i,i+1)|m_i-m_{i+1}| }{\sum_i^{N_M} \omega(i,i+1)}
\end{equation}

\noindent  where $m_i$ and $m_{i+1}$ are consecutive magnitudes in the $M$ filter, in the light curve folded with the trial period, $N_M$ is the number of observations and $\omega(i,i+1)$ are weights proportional to the magnitude errors $\sigma_i$, given by

\begin{equation}
 \omega(i,i+1) = \frac{1}{ \sigma_i^2 + \sigma_{i+1}^2 }
\end{equation}

The true period corresponds to the minimum $S(M)$, since the light curve is smooth when folded with the correct period and therefore the sum of differences between consecutive magnitudes is minimal. Since $S(M)$ considers only the data in the $M$ filter, in order to incorporate the information available for each object in all three $V,R$ and $I$ filters we used the composite index $S_{VRI}$ as the average of the $S(M)$ for each filter \citep[see e.g.][]{Watkins2009}, weighted by the corresponding number of observations, as given by

\begin{equation}\label{e:s_vri}
S_{VRI}=\frac{N_VS(V) + N_RS(R) + N_IS(I)}{N_V+N_R+N_I}
\end{equation}

where $N_V,N_R,N_I$ correspond to the number of observations in each of the $V,R,I$ filters respectively.
By minimizing the $S_{VRI}$ parameter given by Equation \ref{e:s_vri}, we are using the data in all the available filters, requiring that the best period be the one that smoothes the period-folded light curves in the three photometric bands. 

RRLS are known to exhibit a narrow range in temperatures, and therefore colour, which is why the use of colour cuts is frequent in RRLS searches \citep[e.g.][\citetalias{Vivas2004}]{Ivezic2005}. However, in low galatic latitude surveys, stars are severely affected by strong and highly variable extinction, and the use of colour cuts 
is better avoided in order to ensure the survey's completeness, albeit increasing the computational cost of conducting the period search over a much larger sample. Therefore, the period search was conducted over the \emph{full} sample of $486,766$ variable candidates selected without imposing any sort of colour cuts. 
 The search spanned the period range from $0.2-1.2$ d with a $10^{-4}$ d period step and for each star, and a second order search was performed with increased period resolution around the 5 best periods initially found, using a period step of $10^{-5}$ d. The best 3 periods found after the refined search were reported for each object characterized by a statistical significance parameter $\Lambda$ defined as

\begin{equation}\label{e:lambda}
 \Lambda=\frac{S_{VRI}(P) - \langle S_{VRI} \rangle}{\sigma(S_{VRI}) } 
\end{equation}

where $\langle S_{VRI} \rangle$ and $\sigma(S_{VRI})$ are the mean and standard deviation of the distribution of $S_{VRI}$ values for all trial periods. 

\begin{figure*}
\begin{center}
\includegraphics[width=1.7\columnwidth]{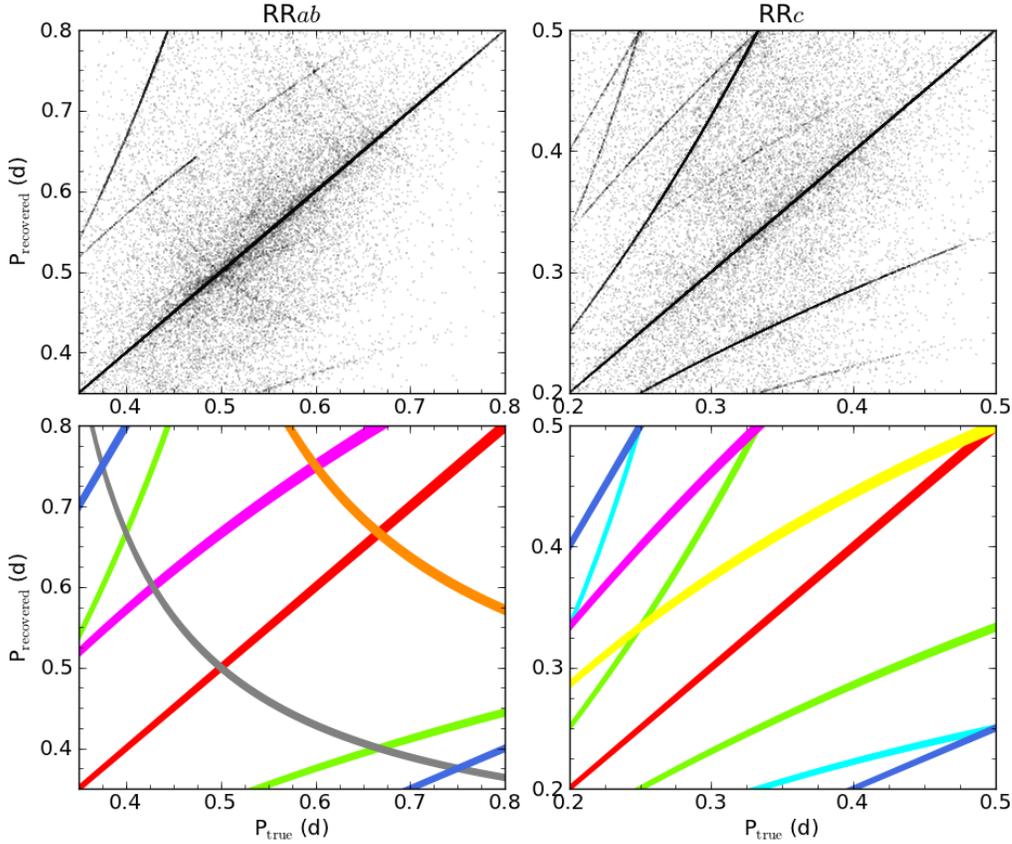}
\caption{\emph{Top:} Recovered versus true period for synthetic stars: \emph{(Left)} \rrab~\emph{(Right)} \rrc. \emph{Bottom:} Loci defined by the period aliases visible in the top panels. The shaded red area indicates the identity locus. The remaining indicate the loci of the most common alias periods: \emph{(Green)} one-day aliases, \emph{(Blue)} $2\times P_{real}$ and $0.5\times P_{real}$,   \emph{(Magenta)} $2\times$ one-day alias, \emph{(Cyan)} $1/2$ day aliases, \emph{(Orange)} $1/3$ day aliases, \emph{(Grey)} $1/4$~day aliases, \emph{(Yellow)} $2\times $ $1/2$-day alias. The shaded areas have a width equal to $2$ per cent separation from the corresponding locus. Note that, in the top panels, the points lying within the identity locus correspond to $80$ per cent and $46$ per cent of the total synthetic \rrab~and \rrc~samples respectively.}
\label{f:aliases_abc}
\end{center}
\end{figure*}

\subsubsection{Period Aliases}\label{s:period_alias}

The emergence of \emph{alias} or \emph{spurious} periods is a common problem when trying to identify the period of a signal from time samplings which may or may not be uniform. In general, these can simply be harmonics of the true period or they can come up as a consequence of an underlying periodicity in the time sampling, being therefore external to the signal itself. Following \citet{Lafler1965} we can express the relationship between the true period $P$ and spurious periods $\Pi$ as

\begin{equation}\label{e:aliasperiods}
 \frac{1}{\Pi} = \frac{k}{P} + \frac{1}{p}
\end{equation}

\noindent where $k$ is a rational number, which produces the true period harmonics; and $p$ is the external period of the sampling. As noted by \citet{Lafler1965} the external periods $p$ of interest are those which produce period aliases in the expected range of the signal, in this case in the period range of RRLSs, since the rest can be ruled out on physical grounds.

In the present case of the RRLS search, the typical one day periodicity between the drift-scan observations composing our survey, leads to the emergence of spurious periods in the range $0.2 \la P({\mathrm d})\la 1$ expected for RRLSs. This spurious periods, commonly termed one-day aliases, are obtained from Equation \ref{e:aliasperiods} with $p=\pm1$ and $k=1$. Other period aliases can arise due to underlying periodicities in the time sampling, as a consequence of the inhomogeneity in the time sampling of our survey.

In order to empiricaly identify the most frequent alias periods in our sample, we conducted the period search described in the previous section, over the synthetic time series data for $\sim130,000$ of our simulated \rrab~and \rrc~stars (see Sec. \ref{s:synth_rrls}. Figure \ref{f:aliases_abc} shows plots of the recovered period as a function of the true period for synthetic \rrab~(left) and \rrc~(right) stars. The most populated sequence in the plot is the identity, indicated by the red shaded area. Different alias periods given by Equation \ref{e:aliasperiods} match the other  sequences observed in Figure \ref{f:aliases_abc}. The most prominent loci defined by the alias periods identified in the sample are shaded in different colours, as detailed in Table \ref{t:alias_freqs}. The width of the shaded loci in Figure \ref{f:aliases_abc} indicates a separation smaller than $2$ per cent from the corresponding main locus.

Figure \ref{f:aliases_abc} shows that the majority of the periods lie in the identity locus. If the recovered period differs in less than $10$ per cent from the true period, we consider it correctly recovered, with this definition we find that for \rrab~$80$ per cent of the times the period is recovered, while for \rrc~stars it is only $\sim46$ per cent. For the majority of the remaining stars the recovered period is an alias of the true period. This indicates, as can be seen in the figure, that the fraction of \rrc~stars affected by spurious periods is much larger than that for \rrab~stars. The frequency of occurence of the alias periods indicated in Figure \ref{f:aliases_abc} is summarized in Table \ref{t:alias_freqs}, for both types of RRLSs, computed from $\sim$130.000 simulated \rrab~and \rrc~light curves. As can be seen both in the table and in Figure \ref{f:aliases_abc}, one-day aliases are the most frequent as noted by \citet{Lafler1965}. Nevertheless, other spurious periods can arise, albeit less frequently, such as the $1/3$-day and $1/4$-day aliases for \rrab~stars and the $1/2$-day alias and first harmonic of the true period for \rrc~stars. 

This empirical overview of the most frequent spurious periods, obtained by means of the simulated light curves, enabled us to quantify the relative importance of the different posible aliases present in our RRLS survey. 
Based on these results, an inspection of the identified period aliases was included in the period search, described in the previous subsection, during the visual inspection of the period-folded light curves. This allowed increasing the completeness in the RRLS identification, which is described in the following section.

\begin{table}
\begin{minipage}{85mm}
\caption{Occurrence frequency of alias periods.}
\begin{scriptsize}
\begin{tabular}{lrrccl}
\hline
Type & $k$ & $p$ & \multicolumn{2}{c}{Frequency (per cent)}  & Color \\
		  &	    &(d)	   &\rrab &\rrc&    \\
\hline
1-d alias               	        & $1$   			& $+1$   		   &  $11.9$  & $13.1$ & Green \\
		               	        & $1$   			& $-1$   		   &  $28.4$  & $36.2$ & Green \\
$1/2             $-d alias 	& $1$  	 		& $+1/2           $  & $ 1.9$   & $1.6$   & Cyan  \\
					& $1$  	 		& $-1/2           $   &   $--$   & $6.1$   & Cyan  \\
$1/3              $-d alias	& $-1$			        & $1/3            $  &  $12.6$  & $--$   & Orange \\
$1/4              $-d alias 	& $-1$			& $1/4            $   &   $17.4$   & $--$   & Grey \\
Harmonics 			& $2$ 			& $\infty$ 	           & $8.4$   & $7.7$ & Blue \\
		 			& $1/2           $ 	& $\infty$ 	          & $1.7$     & $0.4$   & Blue \\
1-d alias 	                        & $1/2           $ 	& $-1/2           $  & $--$  & $15.4$   & Yellow \\
harmonic                   & & & & & \\
$1/2             $-d alias 	& $1/2           $ 	& $-1/4            $  & $17.7$  & $19.4$   & Magenta \\
harmonic                   & & & & & \\
\hline
\end{tabular}
\end{scriptsize}
\label{t:alias_freqs}
\end{minipage}
\end{table}

\begin{figure*}
\begin{center}
\includegraphics[width=1.8\columnwidth]{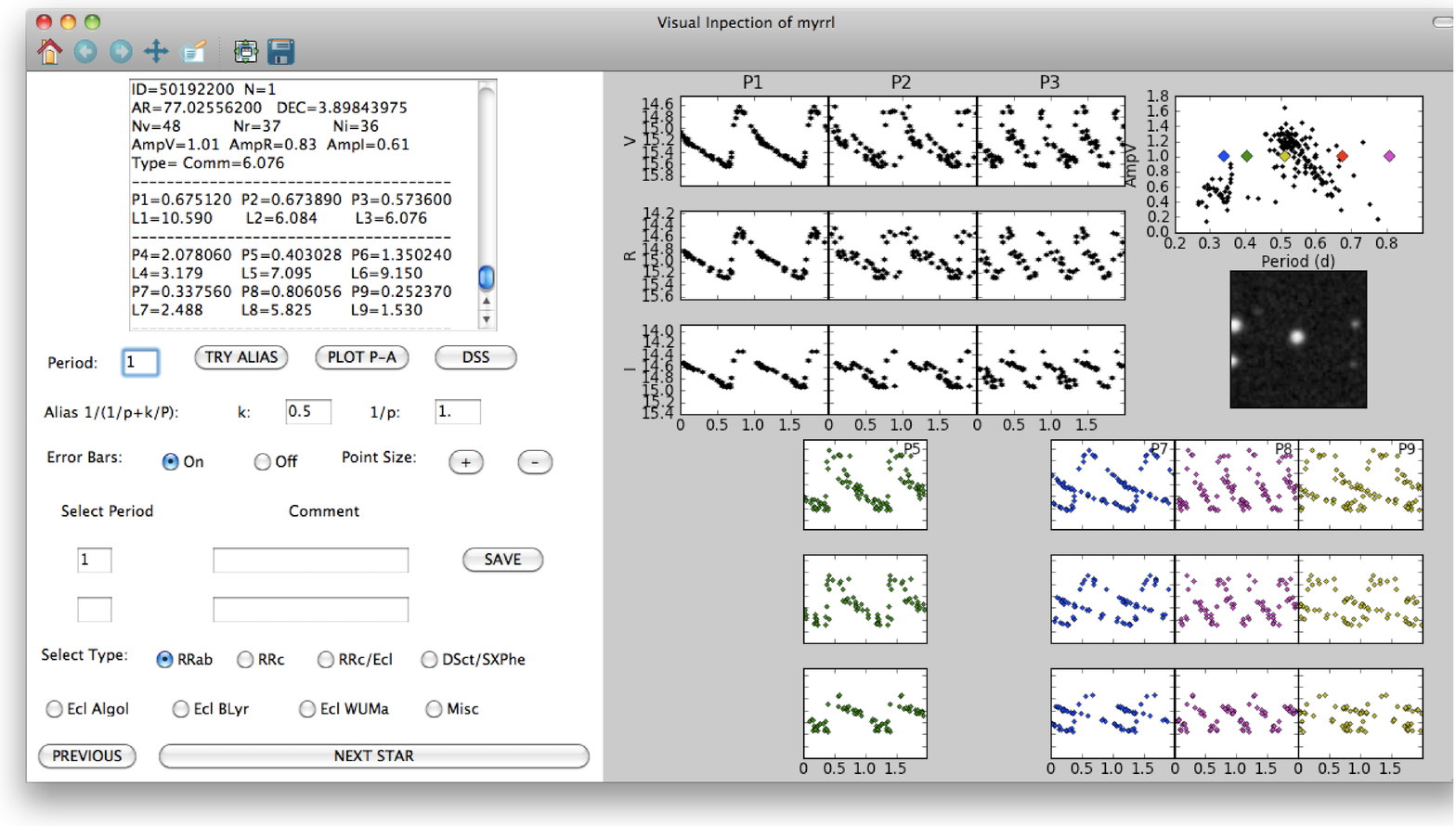}
\caption{Graphical user interface of the \emph{InspectLC.py} tool developed for light curve visual inspection.}
\label{f:inspectlc_gui}
\end{center}
\end{figure*}

\subsubsection{Final Selection of RRLS}

We visually inspected the light curves for the $66,086$ stars for which a statistically significant ($\Lambda\geq3.0$) minimum of $S_{VRI}$ was found. The visual inspection and classification of light curves was done using the plotting tool \emph{InspectLC.py}, developed as part of this work, whose graphical user interface is shown in Figure \ref{f:inspectlc_gui}. The \emph{InspectLC.py} program allows visualizing, for each star, light curves in three different filters (in this case $V,R,I$) period-folded with each of the three most significant periods found. These light curves are shown in Fig. \ref{f:inspectlc_gui} under the labels P1, P2 and P3. The text box to the left of the window summarizes each star's parameters: identification number, $\alpha$, $\delta$, amplitude and number of observations in each of the three filters, the three best periods ($P_i$ with $i=1,2,3$) and their corresponding values of $\Lambda$. The `TRY ALIAS' button displays, for a given period $P_i$, the period-folded light curves using 5 of the most common alias periods (see Table \ref{t:alias_freqs}) plus one user-defined alias, computed from the values inserted in the `$k$' and `$1/p$' textboxes (see Section \ref{s:period_alias} and Equation \ref{e:aliasperiods}). These light curves are displayed in Figure \ref{f:inspectlc_gui} under the labels P4 to P8 and P9 respectively. The program does not display the light curves for alias periods outside the period range of RRLS.  The `PLOT P-A' button plots each period with its corresponding aliases in a period-amplitude plot, as shown in the upper right corner of the window in Figure \ref{f:inspectlc_gui}. This plot also shows data for RRLS in the globular cluster M3 from \citet{Corwin2001} as a reference to illustrate the loci ocuppied by \rrab~and \rrc~stars in this diagram. This plot is useful for deciding between different periods that produce similarly good period-folded light curves. In the above example P1 was selected as the correct period. Aditionally, the `DSS' button allows downloading and plotting a DSS thumbnail image centered on the coordinates of the star, in order to facilitate the identification of contaminants like spurious extended objects, artifacts on very bright stars or double stars. Finally, the `SAVE' button allows saving for each star, in an ascii format file, upto two periods, comments on each period and the variable star classification selected from the bullets on the lower left part of the window.

Finally, a total of 160 \rrab~and 51 \rrc~stars were identified. As a byproduct of  the visual inspection, eclipsing binary candidates were also found and will be discussed in an upcoming paper. The catalogue of $V,R$ and $I$ light curves for all \rrab~and \rrc~stars is shown in Figure \ref{f:rrl_lcs} in Appendix \ref{a:rrl_lcs}. The RRLS light curve parameters, such as amplitudes, maximum-light magnitudes and phase, were computed by fitting light curve templates to the data using \rrab~and \rrc~templates from \citet{Layden1998} and \citet{Vivas2008} respectively, and further refining the period with a step of $10^{-6}$ d. These parameters are listed in Tables \ref{t:rrab_lcparams} and \ref{t:rrc_lcparams} for \rrab~and \rrc~respectively.

In the following sections we describe the effect of period aliases in the catalogue, the completeness estimated in the identification of RRLS, as well as the posible contaminants of the sample.

\begin{table*}
 \begin{minipage}{180mm}
\caption{Light curve parameters for \rrab~stars.}
\begin{tiny}
\begin{tabular}{cccccccccccccccc}
\hline
ID & $\alpha$ & $\delta$ & ($N_V$,$N_R$,$N_I$) & Period & Amp\emph{V} & Amp\emph{R} & Amp\emph{I} & HJD$_{maxl}$ & $\langle V\rangle$ &  $\langle R\rangle$ & $\langle I \rangle$ \\
 & (\degr) & (\degr) &                     &      (d)  &      (mag)        &        (mag)      &         (mag)       &  -2450000 (d) &       (mag)         &         (mag)       &        (mag)\\
\hline
1000&  60.935722 & -1.498512 &(   0,   10,   9)&  0.444400& $\cdots$  &   0.91    &   0.62    & 1200.70430400 &      $\cdots$     & 16.79 $\pm$ 0.02  & 16.55 $\pm$ 0.03 \\
1001&  62.136738 & -2.995186 &(   0,   11,   0)&  0.465461& $\cdots$  &   0.72    & $\cdots$  & 2279.76109714 &      $\cdots$     & 19.16 $\pm$ 0.11  &      $\cdots$    \\
1002&  62.444096 & -4.000918 &(  27,   19,  11)&  0.498114&   1.16    &   1.13    &   0.87    & 1823.74922626 & 17.53 $\pm$ 0.04  & 17.13 $\pm$ 0.02  & 16.49 $\pm$ 0.02 \\
1003&  65.498161 & -3.226088 &(  28,   16,  24)&  0.501271&   1.25    &   1.07    &   0.87    & 1824.19917603 & 15.91 $\pm$ 0.01  & 15.77 $\pm$ 0.02  & 15.40 $\pm$ 0.01 \\
 563&  66.183346 & -3.859660 &(  24,    0,  11)&  0.539828&   0.85    & $\cdots$  &   0.41    & 1867.80606764 & 14.03 $\pm$ 0.02  &      $\cdots$     & 13.60 $\pm$ 0.01 \\
\hline
\end{tabular}
\end{tiny}
\label{t:rrab_lcparams}
\footnotetext{Note: Table \ref{t:rrab_lcparams} is published in its entirety in the electronic edition of the journal. A portion is shown here for guidance regarding its form and content.}
\end{minipage}
\end{table*}

\begin{table*}
 \begin{minipage}{180mm}
\caption{Light curve parameters for \rrc~stars.}
\begin{tiny}
\begin{tabular}{cccccccccccccccc}
\hline
ID & $\alpha$ & $\delta$ & ($N_V$,$N_R$,$N_I$) & Period & Amp\emph{V} & Amp\emph{R} & Amp\emph{I} & HJD$_{maxl}$ & $\langle V\rangle$ &  $\langle R\rangle$ & $\langle I \rangle$ \\
 & (\degr) & (\degr) &                     &      (d)  &      (mag)        &        (mag)      &         (mag)       &  -2450000 (d) &       (mag)         &         (mag)       &        (mag)\\
\hline
1013&  74.151787 & -3.740614 &( 21, 17,  11)&  0.331910&   0.43    &   0.40    &   0.23    & 1824.12209720  & 17.91 $\pm$ 0.05  & 17.67 $\pm$ 0.04  & 17.46 $\pm$ 0.05 \\
1015&  74.725197 &  3.882039 &( 44, 39,  33)&  0.400277&   0.58    &   0.36    &   0.23    & 1870.98566120  & 15.22 $\pm$ 0.01  & 14.83 $\pm$ 0.01  & 14.48 $\pm$ 0.01 \\
1019&  75.680612 &  5.070141 &( 19, 12,  19)&  0.384742&   0.38    &   0.38    &   0.33    & 1871.09802948  & 15.90 $\pm$ 0.01  & 15.77 $\pm$ 0.01  & 15.54 $\pm$ 0.03 \\
1022&  76.733847 &  0.805251 &( 33, 20,   9)&  0.412460&   0.44    &   0.37    &   0.16    & 1539.59914360  & 18.80 $\pm$ 0.07  & 18.63 $\pm$ 0.06  & 18.38 $\pm$ 0.10 \\
1028&  77.485196 &  0.078322 &( 82, 21,  34)&  0.370848&   0.30    &   0.22    &   0.14    & 1539.62626816  & 17.51 $\pm$ 0.04  & 17.40 $\pm$ 0.03  & 17.21 $\pm$ 0.05 \\
\hline
\end{tabular}
\end{tiny}
\label{t:rrc_lcparams}
\footnotetext{Note: Table \ref{t:rrc_lcparams} is published in its entirety in the electronic edition of the journal. A portion is shown here for guidance regarding its form and content.}
\end{minipage}
\end{table*}

\subsubsection{Comparison with Variable Star Catalogues}

The present survey partially overlaps with the RRLS catalogue from \citetalias{Vivas2004} (see Figure \ref{f:survey_coverage_aitoff}), also conducted with QUEST observations but focused on high latitude regions. 
All 21 \rrab~stars and 31 ($\sim78$ per cent) of the \rrc~stars from V04,
expected in our survey's area were succesfully recovered. The remaining $22$ per cent of \rrc~stars were re-classified as contaminants based on their improved light curves which include a larger number of observations, in particular in the $R$ and $I$ bands, with respect to V04. The 52 RRLS recovered from V04 keep their original identification numbers, which correspond to numbers $\leq999$ in Tables \ref{t:rrab_lcparams} and \ref{t:rrc_lcparams}. The 27 RRLS with identification numbers in the range 500-600 are from \citet[in preparation]{Vivas2012}. The $132$ \rrab~and \rrc~stars identified in the present survey have identifiers from $1000$ onwards.

We also cross-matched our RRLS catalogue with the ASAS-3 \citep{Pojmanski2002} and NSVS \citep{Kinemuchi2006}, with a tolerance of $7\arcsec$. Out of the 132 RRLS not present in V04 or \citet{Vivas2012}, 131
are new discoveries: one match is found in NSVS and none in ASAS-3. Our RRLS 1012 corresponds to star 722 of the NSVS. The period reported for this star on the NSVS is $P_{\mathrm{NSVS}}=0.53495$ d which coincides within the errors with our period $P_{\mathrm{QUEST}}=0.53497$ d and in both catalogues, the star was classified as \rrab.

\subsubsection{Completeness}\label{s:completeness}

\begin{figure*}
\begin{center}
\includegraphics[width=1\textwidth]{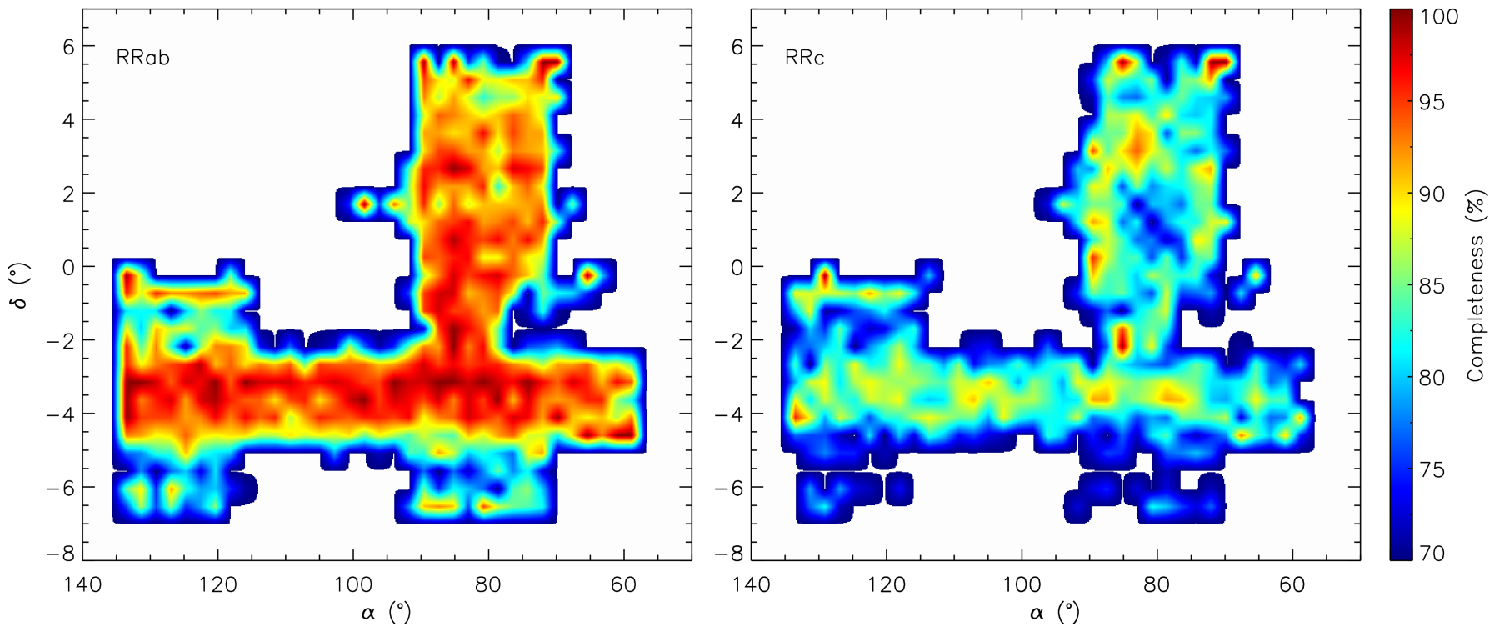}
\caption{Map of average completeness in the identification of \rrab~(\emph{left}) and \rrc~(\emph{right}) stars, computed from synthetic RRLS within the photometric completeness limits of the  survey ($14.0 \leqslant V\leqslant18.5$).}
\label{f:completeness_rrab_rrc}
\end{center}
\end{figure*}

The completeness achieved in the RRLS identification was estimated from the catalogue of synthetic light curves for \rrab~and \rrc~stars described in Sec. \ref{s:synth_rrls}. The completeness was computed, independently for each RRLS type, as the percentage of synthetic stars recovered after the period search described in Sec. \ref{s:period_search} was performed on a sample of $\sim65,000$ synthetic RRLS of each type. A synthetic star was considered succesfully recovered if one of the 3 best periods found were within $10$ per cent of the true light curve period or one of the period aliases described in Sec. \ref{s:period_alias}. The average completeness was computed for synthetic stars in the magnitude range $14.0 \leqslant V\leqslant18.5$ as a function of position in the survey area, since it strongly depends on the time sampling and number of observations in each filter. 

The resulting completeness maps for RRLS of types \typeab~(left) and \typec~(right) are shown in Figure \ref{f:completeness_rrab_rrc} in equatorial coordinates. As illustrated by the figure, for \rrab~stars the average completeness is $\sim95$ per cent for most of the survey area, except in the declination stripe $-6\degr \leqslant \delta \leqslant -4\degr$, where the number of observations per object is typically less than $15$ in all three photometric bands (see Figure \ref{f:map_nvri}). For \rrc~stars the estimated completeness is $\sim80-85$ per cent, significantly lower than for \rrab~stars, as expected since \rrc~light curves have smaller amplitudes.

\subsubsection{Possible Contaminants}\label{s:contams}

The main source of contamination in an RRLS survey are eclipsing binaries (mostly of the WUMa type), $\delta$ Scuti and SX Phoenicis stars. Eclipsing contact binaries or WUMa stars, consist of a system in which both stars have filled their Roche lobes and are undergoing mass transfer \citep{Sterken2005} and have light curves showing two eclipses of very similar depths. $\delta$ Scuti stars are pulsating main sequence dwarfs and blue stragglers, while SX Phoenicis stars are their metal-poor counterparts; both types are also called Dwarf Cepheids \citep{Nemec1990,Sandage2006} and have asymmetric light curves, similar to those of RRLSs, though with smaller periods ($P<0.2$ d) and amplitudes \citep[$\mathrm{Amp}V\la0.2$ mag,][]{Sterken2005}. 

\begin{figure}
\begin{center}
\includegraphics[width=0.93\columnwidth]{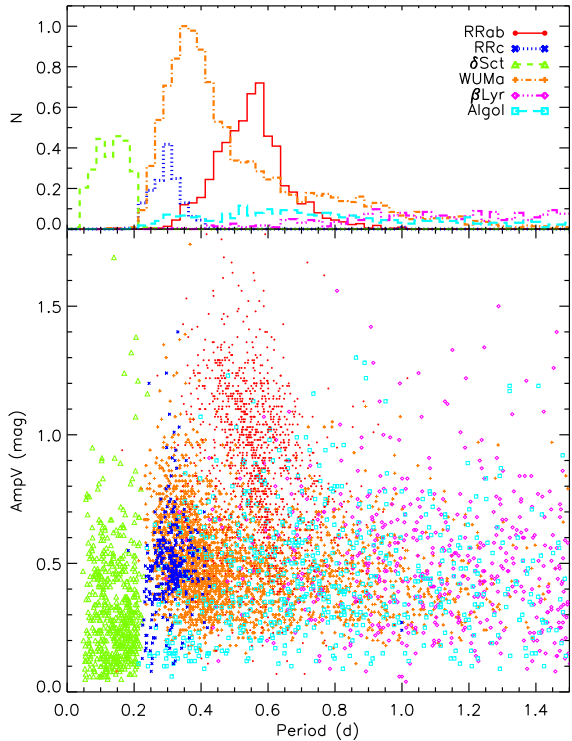}
\caption{\emph{Bottom:} $V$ amplitude versus period for variable stars of type \rrab, \rrc, $\delta$ Scuti, Algol, $\beta$ Lyrae and WUMa from the ASAS-3 catalogue \citep{Pojmanski2002}. \emph{Top:} Period histogram for the different variable star types illustrated in the bottom panel.}
\label{f:P_Amp_contaminants}
\end{center}
\end{figure}

A plot of $V$ amplitude versus period is shown in Figure \ref{f:P_Amp_contaminants} for \rrab, \rrc, Algol, $\beta$ Lyrae, WUMa and $\delta$ Scuti stars from the ASAS-3 survey \citep{Pojmanski2002}. As illustrated by the figure, the period and amplitude ranges for most of these stars largely overlap, making the light curve shape an important criteria in distinguishing between the different types of variables. Most of the contamination problems affect the \rrc~rather than the \rrab~stars.

Eclipsing binary light curves, of any type, are unlikely to be mistaken for RRLSs of type \typeab, while dwarf cepheids, though having similar light curves, have periods and amplitudes which are too small for \rrab~stars. A few dwarf cepheids could have alias periods in the range of RRLS, however in this case the contaminants are unlikely to  follow the well known Period-Amplitude relationship for \rrab~stars \citep[e.g.][]{Smith1995,Cacciari2005}, which can in turn be used as a criterion to discard them.  This criterion is also helpful to avoid other possible contaminants of the \rrab~sample, such as Anomalous Cepheids and short period Type-II Cepheids \citep{Sandage2006}, also known as BL Herculis and W Virginis stars, which have light curve shapes similar to \rrab~stars with periods in the range from $1-50$ d (W Vir)  and $1-7$ d (BL Her), partially overlapping the range for \rrab~stars in the long-period end. Therefore, the contamination expected in the \rrab~sample can be considered negligible.

The \rrc~sample is, on the contrary, more prone to contamination. 
The light curves of eclipsing binaries of the WUMa and $\beta$ Lyrae type can resemble those of \rrc~stars if period-folded with half the true period of the binary. However, as noted by \citet{Kinman2010}, an important difference between \rrc~stars and eclipsing binaries is that pulsating stars have larger amplitudes in bluer bands compared to red bands; while eclipsing binaries have similar amplitudes in different photometric bands \citep{Kinman2010,Sterken2005}. This is a useful criterion for differentiating both types of variables, except for variable stars with poorly sampled light curves or very small amplitudes ($<0.2-0.3$ mag), in which case the distinction is more difficult. On the other hand, since this behaviour of the amplitude in different photometric bands is common to pulsating stars, it is not useful for distinguishing \rrc~stars from the larger amplitude $\delta$ Scuti or SX Phoenicis stars, making these likely contaminants of the sample.

Therefore, the contamination expected in the \rrc~sample could be significant. In particular, among the \rrc, a large number of WUMa and $\delta$ Scuti stars belonging to the thin disk is expected given the low latitudes being surveyed and the fact that we are not impossing any colour cuts which would sizeably have reduced the contamination \citep[\citetalias{Vivas2004}]{Ivezic2000}. This added to the relatively low completeness of the \rrc~sample ($\sim80$ per cent), makes it inappropriate to use the \rrc~stars in the computation of the thick disk structural parameters, which is why we have restricted the following analysis to the \rrab~sample.

\subsection{Physical Parameters of RRL stars}\label{s:rrls_phys}

\subsubsection{Extinctions}\label{s:exts}

The present survey spans areas at very low galactic latitudes ($|b|<15\degr$), where the extinction can be very high and spatially variable. The \citet*{Schlegel1998} dust maps, although covering the whole sky, are unreliable at galactic latitudes lower than $\sim10\degr$. On the other hand, a correct estimate of the extinction is crucial in the determination of distances for the survey RRLSs and their subsequent use in the computation of density profiles.

In addition to being excellent standard candles, RRLSs can also be used as colour-standards. \citet{Sturch1966} finds the average $\BV$ colour index during the phase of minimum light ($\phi\in[0.5,0.8]$) is approximately constant for \rrab~stars, having a small dispersion of $\sim0.03$ mag; which makes it possible to estimate the extinction affecting a particular \rrab~star by simply comparing the observed minimum-light colour with the intrinsic colour. Later on, the calibration was extended to $\VI$~by \citet{Day2002} and \citet{Guldenschuh2005}, and to $\VR$ by \citet*{Kunder2010}; obtaining smaller standar deviations of $\sim0.02$ mag in these colour indices.

In the present work, we used the minimum-light colour calibrations $(\VR)_{min}^{0.5-0.8}= 0.27 \pm 0.02$ from \citet{Kunder2010} and $(\VI)_{min}^{0.5-0.8}= 0.58 \pm 0.02$ from \citet{Guldenschuh2005}. For the $\sim40$ \rrab~stars that do not have $V$-band observations we used the intrinsic colour $(\RI)_{min}^{0.5-0.8}= 0.31 \pm 0.03$, computed from the previous two colours. The observed minimum-light colours were computed from the subtraction of the best fitting templates in the two bands used. Comparing these with the assumed intrinsic colours we computed the corresponding $E(\VR)$, $E(\VI)$ and $E(\RI)$ colour excesses and extinctions $A_V,A_R$ y $A_I$  were derived assuming the standard reddenning law from \citet*{Cardelli1988}.

We used the 56 stars with observations in all three filters, to compute the residuals in the $E(B-V)$ derived using the $\VR$, $\VI$ and $\RI$ colour indices, in order to test the consistency of the resulting extinctions. The mean differences obtained were $E(B-V)_{V\!R}-E(B-V)_{V\!I}=-0.007$ mag and $E(B-V)_{R\!I}-E(B-V)_{V\!I}=-0.006$ mag, with standard deviations of $0.092$ mag and $0.082$ mag respectively, showing there is very good agreement between the different measurements. For stars with data in the $VRI$ filters the adopted $E(B-V)$ were computed as the error-weighted mean of the colour excesses computed from the $\VR$ and $\VI$~intrinsic colours and the corresponding errors were computed by standard error propagation.

\begin{figure}
\begin{center}
\includegraphics[width=1.\columnwidth]{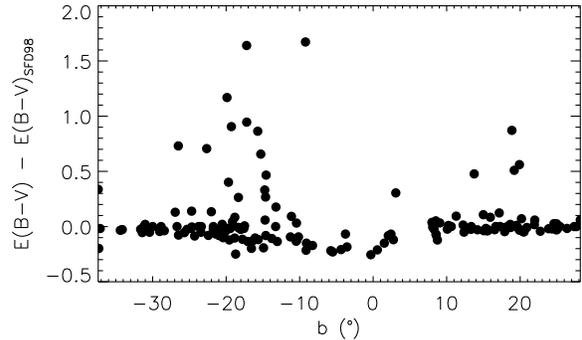}
\caption{Residuals between colour excesses $E(B-V)$ from \citet{Schlegel1998} dust maps and those computed from \rrab~intrinsic colours, as a function of galactic latitude $b$.}
\label{f:ebv_vs_b}
\end{center}
\end{figure}

Residuals between these colour excesses and those obtained from interpolations in the \citet{Schlegel1998} dust maps are shown in Figure \ref{f:ebv_vs_b}. The plot shows the residuals are nearly zero for moderately high galactic latitudes ($|b|\geq20\degr$), while as the latitude decreases the colour excesses computed from \rrab~intrinsic colours are systematically smaller than those from \citet{Schlegel1998}. This behaviour is expected since the former correspond to the colour excesses integrated up to the distance of each RRLS, while the latter are integrated along the entire line of sight. Figure \ref{f:ebv_vs_b} also shows larger residuals around latitudes $b\sim-20\degr$ and $b\sim+10\degr$, which are presumably due to abrupt variations in the extinction on relatively small scales, mainly due to the presence of the Orion Molecular Cloud (at $b\sim-20\degr$). The points showing the larger residuals in Figure \ref{f:ebv_vs_b} coincide with regions where the extinction changes  abruptly and the colour excesses derived from \rrab~stars are systematically larger than those reported by \citet{Schlegel1998}, which is to be expected since the \citeauthor{Schlegel1998} maps have relatively large pixels with $6.1\arcmin\times6.1\arcmin$ and abrupt variations on smaller scales will be averaged out, while \rrab~stars probe the extinction exactly on the line of sight towards each star.

\begin{figure*}
\begin{center}
\includegraphics[width=1.8\columnwidth]{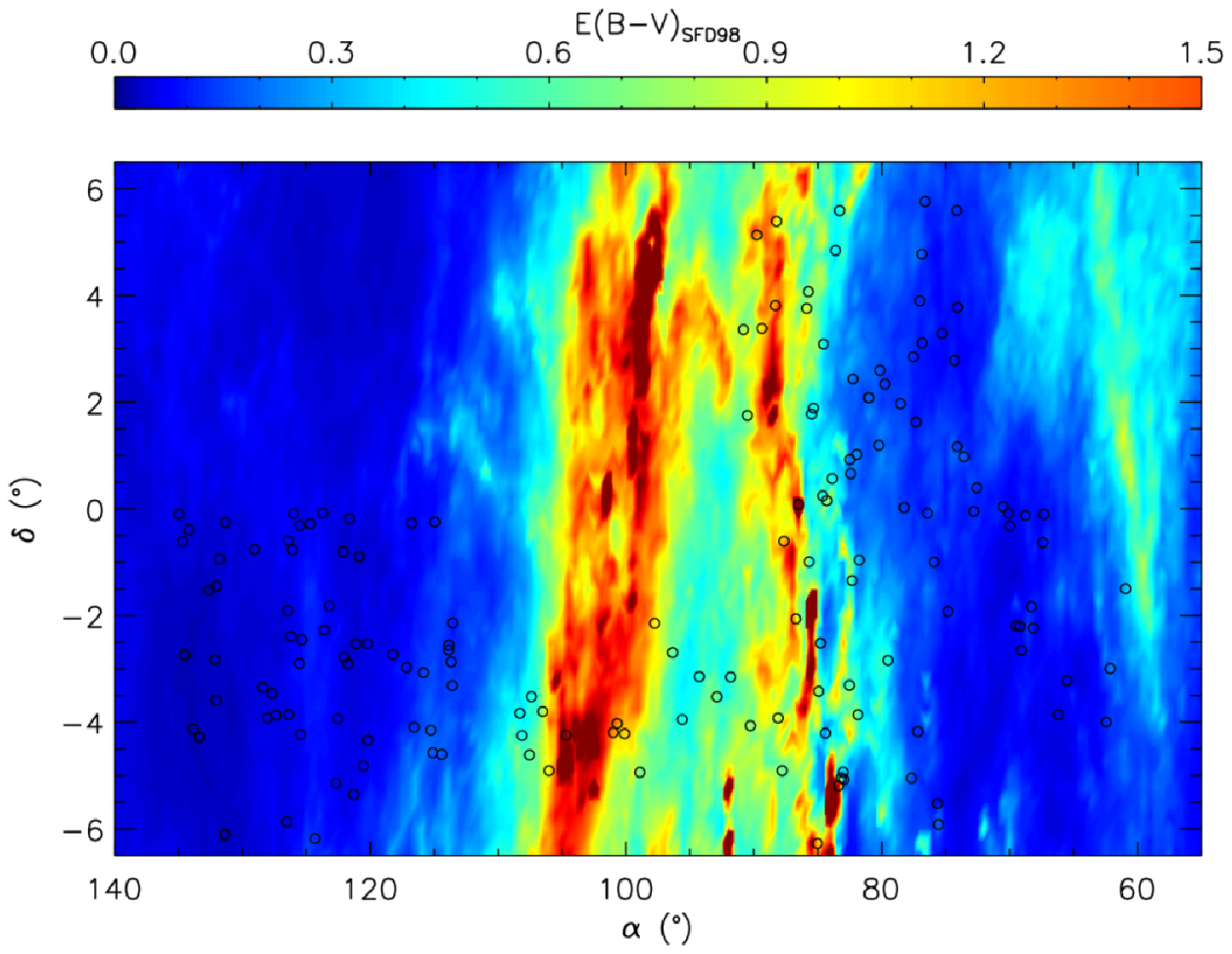}
\caption{$E(B-V)$ map from \citet{Schlegel1998}. Open circles indicate survey \rrab~stars.}
\label{f:map_ebv}
\end{center}
\end{figure*}

\subsubsection{Photometric Metallicities}\label{s:FeH}

RRLSs of type \typeab~exhibit a well known relationship between metallicity $\FeH$, pulsation period $P$ and the $\phi_{31}$ phase of a Fourier light curve decomposition, discovered by \citet{Jurcsik1996}. Also, \citet{Sandage2004} finds that $\phi_{31}$ is correlated with the light curve amplitude and finds an expression analogous to that of \citet{Jurcsik1996} relating $\FeH$ with the light curve amplitude and period for \rrab~stars.

\paragraph{Jurcsik \& Kovacs Metallicities}

\citet{Jurcsik1996} found an empiric relationship between the metallicity $\FeH$, period $P$ and $\phi_{31}$ phase of a Fourier decomposition of \rrab~light curves, expressed in their Equation 3 as

\begin{equation}\label{e:jk_feh}
 \FeH = -5.038 - 5.394P + 1.345\phi_{31}
\end{equation}

The dispersion in this relationship was estimated by \citet{Jurcsik1996} to be $0.11$ dex from a comparison with spectroscopically determined metallicities.

The Fourier fitting process is very sensitive to the light curve sampling. For relatively well sampled curves, but with a moderate number of observations ($N\sim30$), the fits usually have excessive ripples which translate into undesirable uncertainties in the determination of $\phi_{31}$. In order to circumvent this problem, we used the method proposed by \citet{Kovacs2007} which consists in fitting an observed light curve with one of their $248$ light curve templates, representative of \rrab~stars, and determining $\phi_{31}$ from the Fourier decomposition of the best-fitting template\footnote{We used the \emph{tff.f} software from \citet{Kovacs2007} in the template fitting and Fourier decomposition processes}. This allows for robust Fourier decompositions of light curves with a moderate number of observations ($N\ga20$). In order to have an estimate of the quality of the fit, we use the $D_F$ index defined by \citet{Jurcsik1996}, which tests for consistency of the Fourier decomposition parameters using the fact that different Fourier phases $\phi_{ij}$ are correlated (see their Table 6). The restriction $D_F<3$ was applied in order to ensure the reliability of the Fourier fits used in the computation of metallicities, as suggested by \citet{Jurcsik1996}.

\paragraph{Sandage Metallicities} 

Using a sample of \rrab~stars having spectroscopically derived metallicities, \citet{Sandage2004} finds the following empirical relationship between $\FeH$ and the light curve period, amplitude

\begin{multline}\label{e:sandageFeH}
\mathrm{[Fe/H]}_V=(-1.45\pm 0.04)\mathrm{Amp}_V - (7.99 \pm 0.09) \log P \\- (2.145 \pm 0.025)
\end{multline}

\citeauthor{Sandage2004} argues this relationship reflects the same behaviour found by \citet{Jurcsik1996},
since $\phi_{31}$ correlates directly with the light curve amplitude. In practice, \citeauthor{Sandage2004}'s
relation can be used in many more RRLS catalogues since it only requires an accurate knowledge of period and amplitude, while in order to perform the Fourier fits of \citet{Jurcsik1996} the full time series data are required and the light curves must have an appropriate time sampling. 

The relationship expressed in Equation \ref{e:sandageFeH} is calibrated for the $V$ band, found in the majority of RRLS observations in the literature. In our catalogue of 160 \rrab~stars, $25$ per cent of the stars have observations only in the $R$ and/or $I$ bands and not in $V$, which made it necessary to extend Sandage's relation for these photometric bands. Using the photometric metallicities computed from Equation \ref{e:sandageFeH} for $49$ of our stars which are well-observed in $V$ and $R$ ($N_V\geq20$ and $N_R\geq20$) and $45$ stars in $V$ and $I$ ($N_V\geq20$ and $N_I\geq20$), we obtained analogous equations for the $R$ and $I$ bands, via least squares fitting of the corresponding metallicities, periods and amplitudes. The corresponding equations are 

\begin{multline}\label{e:sandageFeHR}
\FeH_R  =  (-1.052 \pm 0.027)\mathrm{Amp}_R \\
- (6.281 \pm 0.087) \log P - (2.254 \pm 0.028)
\end{multline}
\begin{multline}\label{e:sandageFeHI}
\FeH_I  =  (-1.469 \pm 0.032)\mathrm{Amp}_I \\
- (6.516 \pm 0.091) \log P -  (2.250 \pm 0.027)
\end{multline}

\begin{figure}
 \begin{center}
\includegraphics[width=\columnwidth,trim=16mm 4mm 6mm 8mm, clip]{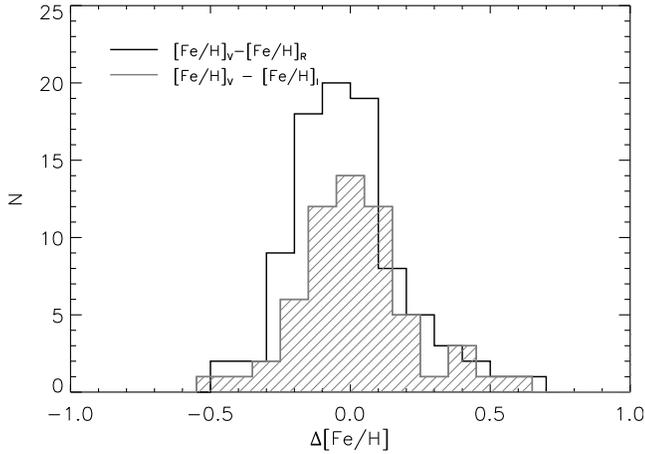}
 \caption{Distribution of differences for photometric metallities [Fe/H]$_R$ and [Fe/H]$_I$ computed from $R$ and $I$ band amplitudes (Equation \ref{e:sandageFeHR} and \ref{e:sandageFeHI}), with respect to [Fe/H]$_V$, computed from the $V$ band amplitude (Equation \ref{e:sandageFeH}).} 
\label{f:hist_FeHresiduos}
\end{center}
\end{figure}

Figure \ref{f:hist_FeHresiduos} shows the distribution of residuals of photometric metallicities $\FeH_R$ and $\FeH_I$ computed from the $R$ and $I$ band amplitudes and Equation \ref{e:sandageFeHR} and \ref{e:sandageFeHI} respectively, with respect to $\FeH_V$ derived from Equation \ref{e:sandageFeH} \citep{Sandage2004}. The two distributions show the agreement between the photometric metallicities computed from Equation \ref{e:sandageFeHR} and \ref{e:sandageFeHI} derived in this work, and those computed via Equation \ref{e:sandageFeH} from \citet{Sandage2004}, resulting in mean residuals of $-0.01$ dex and $0.006$ dex respectively. The intrinsic dispersion of photometric metallicities computed from Equation \ref{e:sandageFeH}, in comparison with spectroscopic metallicities, was estimated by \citet{Sandage2004} as $\sigma_{\FeH}^V=0.26$. Standard deviations for the residual distributions in Figure \ref{f:hist_FeHresiduos} were found to be $0.19$ dex and $0.17$ dex for metallicities computed from $R$ and $I$ band amplitudes respectively. Adding these dispersions in quadrature with $\sigma_{\FeH}^V$, we estimated the intrinsic dispersion of Equations \ref{e:sandageFeHR} and \ref{e:sandageFeHI} to be $\sigma_{\FeH}^R=0.32$ dex and $\sigma_{\FeH}^I=0.31$ dex respectively. Uncertainties in the reported photometric metallicities were computed by adding in quadrature the errors obtained from error propagation in Equations \ref{e:sandageFeH}-\ref{e:sandageFeHI} with the respective intrinsic dispersion of each equation. Using these procedures, the first term accounts for the amplitude uncertainty (the contribution of period uncertainties are negligible) while the second term accounts for the intrinsic dispersion of the method. The typical uncertainties obtained for the photometric metallicities are $\sim0.4$ dex.

\begin{figure}
\begin{center}
\includegraphics[width=\columnwidth]{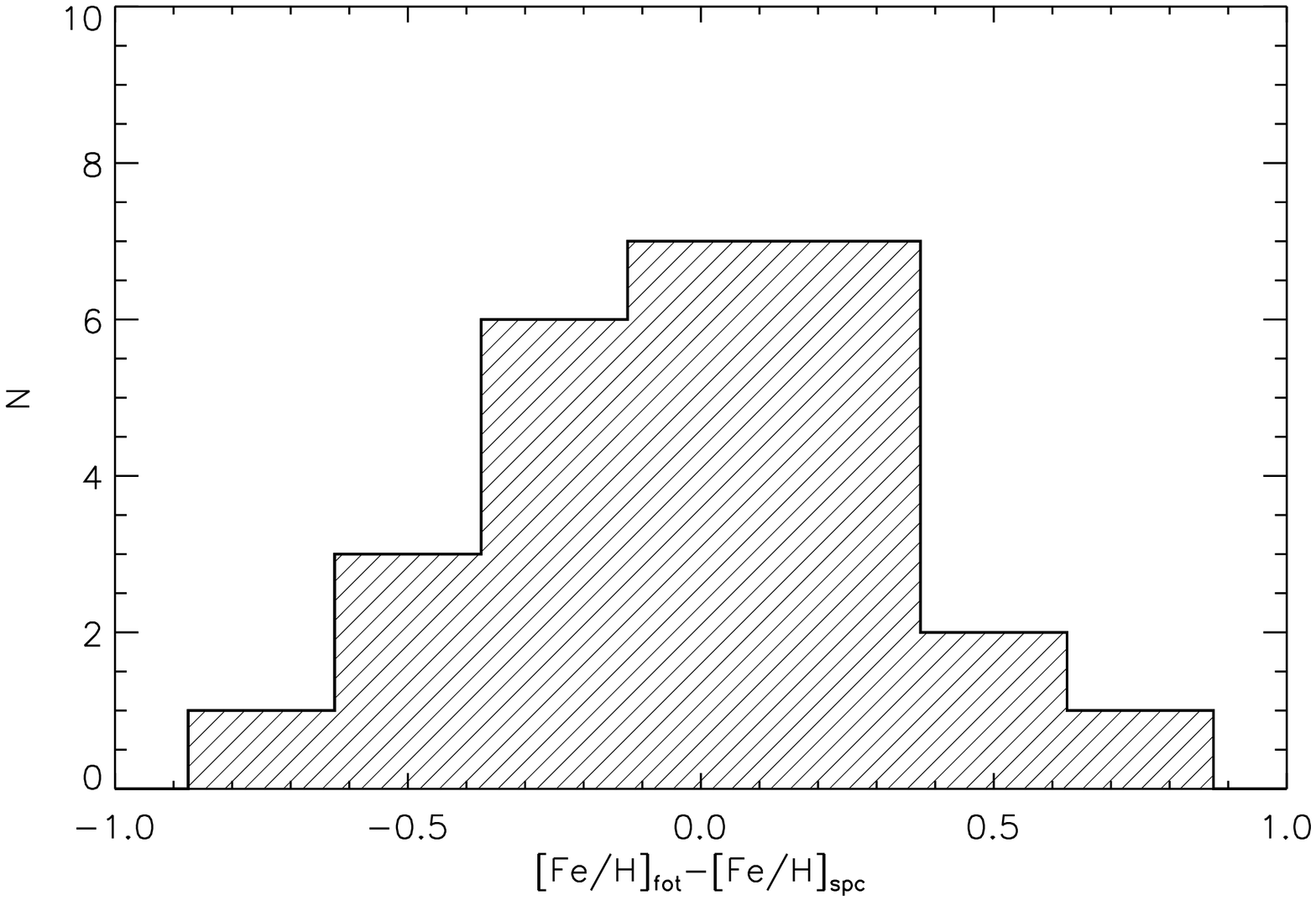}
 \caption{Distribution of residuals of photometric $\FeH_{fot}$ and spectroscopic $\FeH_{spc}$ metallicities for 27 \rrab~in the \citet{Vivas2008} catalogue.}
\label{f:hist_FeHresiduos_fotspec}
\end{center}
\end{figure}

An additional test on these methods was performed by comparing the spectroscopic metallicities for $27$ \rrab~stars from the \citet{Vivas2008} catalogue, with the corresponding photometric metallicities computed from Equation \ref{e:sandageFeH}. Figure \ref{f:hist_FeHresiduos_fotspec} shows the resulting distribution of residuals $\FeH_{fot}-\FeH_{spc}$. The mean and standard deviation of these residuals are $-0.01$ dex and $0.34$ dex respectively, in agreement with the intrinsic dispersions for Equation \ref{e:sandageFeH}, \ref{e:sandageFeHR} and \ref{e:sandageFeHI}.

\paragraph{Metallicity Distribution} 

The photometric metallicities adopted for \rrab~stars were calculated from a weighted average of the metallicities computed using the \citet{Jurcsik1996} (Equation\ref{e:jk_feh}) and \citet{Sandage2004} (Equation \ref{e:sandageFeH}-\ref{e:sandageFeHI}) methods. For those stars for which the Fourier decomposition obtained was unreliable (i.e. $D_F>3$), the metallicity computed from \citeauthor{Sandage2004}'s method was adopted. The metallicities reported here are on the \citet{Zinn1984} scale, and for all stars lie inside the metallicity range ($-2.5\leqslant\FeH\leqslant+0.07$) of the stars used in the calibration of the \citet{Jurcsik1996} and \citet{Sandage2004} methods. 

\begin{figure}
\begin{center}
\includegraphics[width=\columnwidth]{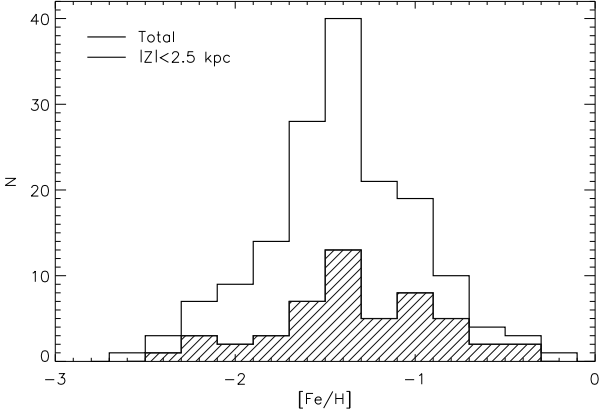}
 \caption{Photometric metallicity distribution for the 160 \rrab~stars in the survey (black). The filled histogram (blue) shows the distribution of metallicities for stars near the Galactic Plane with $|z|\leqslant2.5$ kpc.}
\label{f:hist_FeH}
\end{center}
\end{figure}

The distribution of metallities obtained is shown in Figure \ref{f:hist_FeH}, for the full \rrab~sample (black) and for the stars close to the Galactic Plane with $|z|\leqslant2.5$ kpc. The total metallicity distribution extends from $\FeH\sim-2.5$ up to $\FeH\sim+0.5$ with a mode at $\FeH\sim-1.48$ as can be seen in the figure. The main peak coincides with the expected peak of the halo metallicity distribution, which has an observed metallicity $\FeH=-1.5$ according to \citet{Carney1994} and \citet{Layden1995}, for F and G field halo dwarfs and RRLs respectively. The metallicity distribution near the Galactic Plane shows a hint of an excess for $\FeH\gsim-1.0$, around the metallicity expected for thick disk stars around $\FeH\sim-0.8$ \citep{Gilmore1995,Layden1995}. 

\subsubsection{Distances}\label{s:distances}

The computation of heliocentric distances for the \rrab~stars was done using absolute magnitudes $M^{RRL}_V$ and $M^{RRL}_I$ derived from the Period-Luminosity-Metallicity relations in \citet*[their Equation 8 and 3 respectively]{Catelan2004}. According to \citet{Catelan2004} the use of Period-Luminosity-Metallicity relations is preferable over $M_V-\FeH$ relations, since the former allow accounting for evolution off the Zero-Age Horizontal Branch (ZAHB). The total metal abundance $Z$ was computed from $\FeH$ using Equation 10 in \citet{Catelan2004}, assuming an $\alpha$ to Fe abundance of [$\alpha$/Fe]=$0.2$ dex typical of halo and thick disk stars \citep[e.g.][]{Venn2004,Reddy2003}. For the 9 \rrab~stars which only have $R$ band photometry, we computed $M^{RRL}_R$ by interpolating the corresponding coefficients in Table 4 of \citet{Catelan2004}. 

Finally, the heliocentric distances were computed from Pogson's equation $\log R_{hel}=(m_i-A_i-M_i+5)/5$, where  $A_i$, $M_i$ and $m_i$ correspond to the extinction, observed and absolute magnitudes in the $i$ filter. The corresponding errors were computed via error propagation in this equation, resulting in typical distance errors of $\sim7$ per cent. The resulting heliocentric distance distribution is shown in Figure \ref{f:rhel_hist} and spans the range $0.2 \leqslant R_{hel} ({\rm kpc})\leqslant80$. In Table \ref{t:rrab_physparams} we summarize the colour excesses, metallicities, heliocentric distances $R_{hel}$, as well as height above the Plane $z$ and radial distance $R$ projected on the Galactic Plane, obtained for all \rrab~stars. 

 \begin{figure}
 \begin{center}
\includegraphics[width=\columnwidth]{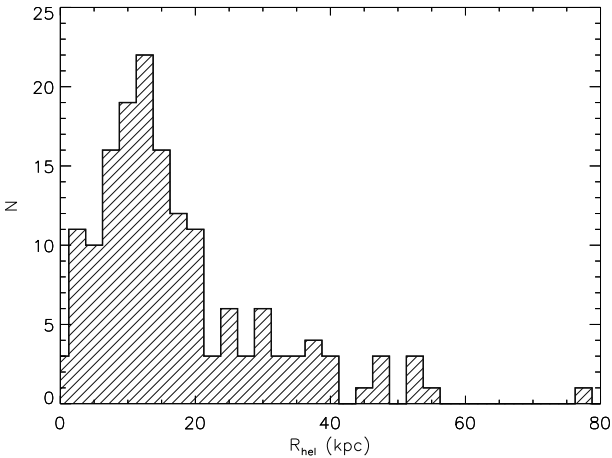}
 \caption{Distribution of heliocentric distances of survey \rrab~stars.} \label{f:rhel_hist}
\end{center}
\end{figure}

\begin{table*}
 \begin{minipage}{90mm}
\caption{Distance, colour excess and metallicities for \rrab~stars.}
\begin{footnotesize}
\begin{tabular}{cccccc}
\hline
  ID & $E(B-V)$ & $\FeH$ & $R_{hel}$   &  $z$        &  $R$ \\ 
& (mag) & (dex) & (kpc)   &  (kpc)        &  (kpc) \\
\hline
 1000 &   0.02 &   -1.0 $\pm$  0.3 &   17.1 $\pm$  1.3 &  -10.4 $\pm$  0.8 &   21.5 $\pm$  1.0 \\
 1001 &   0.13 &   -0.9 $\pm$  0.3 &   46.4 $\pm$  2.5 &  -28.0 $\pm$  1.5 &   44.8 $\pm$  1.9 \\
 1002 &   0.41 &   -1.4 $\pm$  0.5 &   13.0 $\pm$  1.7 &   -7.9 $\pm$  1.0 &   18.1 $\pm$  1.3 \\
 1003 &   0.01 &   -1.6 $\pm$  0.1 &   11.5 $\pm$  0.8 &   -6.5 $\pm$  0.5 &   17.3 $\pm$  0.6 \\
  563 &   0.01 &   -2.1 $\pm$  0.1 &    5.0 $\pm$  0.2 &   -2.8 $\pm$  0.1 &   12.0 $\pm$  0.2 \\
\hline
\end{tabular}
\end{footnotesize}
\label{t:rrab_physparams}
\footnotetext{Note: Table \ref{t:rrab_physparams} is published in its entirety in the electronic edition of the journal. A portion is shown here for guidance regarding its form and content.}
\end{minipage}
\end{table*}

\section{SUMMARY} \label{s:conclusions}

We have presented the QUEST RR Lyrae survey at low galactic latitude, which spans an area of $476$ deg$^2$ on the sky, with multi-epoch observations in the $V$, $R$, and $I$ photometric bands for $6.5\times10^6$ stars in the galactic latitude range $-30\degr \leqslant b \leqslant +25\degr$. The survey has a mean number of $30$ observations per object in each of the $V$ and $I$ bands and $\sim25$ in the $R$ band, while in the best sampled areas each object can have up to $\sim120-150$ epochs in $V$ and $I$ and up to $\sim100$ in $R$. The completeness magnitudes of the survey are $V=18.5$ mag, $R=18.5$ mag and $I=18.0$ mag with saturations of $V=R=14.0$ mag and $I=13.5$ mag.

The survey has identified 211 RRLSs, 160 \textit{bona fide} stars of type \typeab~and 51 strong candidates of type~\typec. The RRLS catalogue presented here contains the positions, mean magnitudes in $VRI$, periods, amplitudes, photometric metallicities, distances and individual extinctions computed from minimum light colours for each star, as well as the period-folded light curves for \rrab~stars (Tables \ref{t:rrab_lcparams}, \ref{t:rrc_lcparams}, \ref{t:rrab_physparams} and Figure \ref{f:rrl_lcs}). The typical distance errors obtained were $\sim7$ per cent. The completeness of the RRLS survey was estimated in $\ga95$ per cent for \rrab~ and $\sim85$ per cent for \rrc~stars, and was computed from light curve simulations reproducing the time sampling and other observing characteristics of our survey, as well as the variable identification and period searching techniques used. 

Our RRLS survey spans \emph{simultaneously} a large range of heliocentric distances $0.5 \leqslant R_{hel}\mathrm{(kpc)} \leqslant 40$ and heights above the plane $-15\leqslant z\mathrm{(kpc)} \leqslant +20$,
with well characterized completeness across the survey area. Combining ours with a bright RRLS survey \citep[such as  those from ][see Sec. \ref{s:intro}]{Layden1995,Maintz2005} to increase the distance coverage even further, will result in a very good tool for studying the Galactic thick disk's structure, in particular for the determination of the density profile's scale length, scale height as well as their dependence with radial distance in order to explore the flare and the truncation profile of the thick disk. This way, the QUEST RRLS from this work will help put constraints on the structure of the more external parts of the thick disk toward the Anticenter, as well as the mechanisms that could have contributed to the formation of the Galactic thick disk. These issues will be addressed in detail in an upcoming paper of the series.

\section*{Acknowledgements}

This research was based on observations collected at the J\"urgen Stock 1m Schmidt telescope of the National Observatory of Llano del Hato Venezuela (NOV), which is operated by CIDA
for the Ministerio del Poder Popular para Ciencia y Tecnolog{\'\i}a, Venezuela. 
The facilities of the 0.9m of the SMARTS Consortium and the YALO telescope at CTIO, 
Chile were also used. C.M. is pleased to thank Gladis Magris, Gustavo Bruzual and Carlos Abad 
for stimulating and helpful discussions. 
C.M. acknowledges support from doctoral grants of the 
Academia Nacional de Ciencias F\'{\i}sicas, Matem\'aticas y Naturales of Venezuela
and CIDA. The work reported here was supported in part by grant S1-2001001144 from FONACIT, Venezuela. 
R.Z. acknowledges the support of NSF grant AST-1108948 from the U.S. government.
The authors are grateful for the assistance of the personnel, service-mode observers, 
telescope operators and technical staff at CIDA and CTIO, who made possible the acquisition of
photometric observations at the NOV and SMARTS telescopes.
The	software package TOPCAT (http://www.starlink.ac.uk/topcat/) was used extensively 
in the preparation of this paper.

\appendix
\section{RRLS LIGHT CURVES}\label{a:rrl_lcs}

The catalogue of $V,R$ and $I$ period-folded light curves for all \rrab~and \rrc~stars found by the survey, is shown in Figure \ref{f:rrl_lcs}1. 

\setcounter{figure}{1}
\begin{figure*}
\begin{center}
 \includegraphics[width=0.9\textwidth]{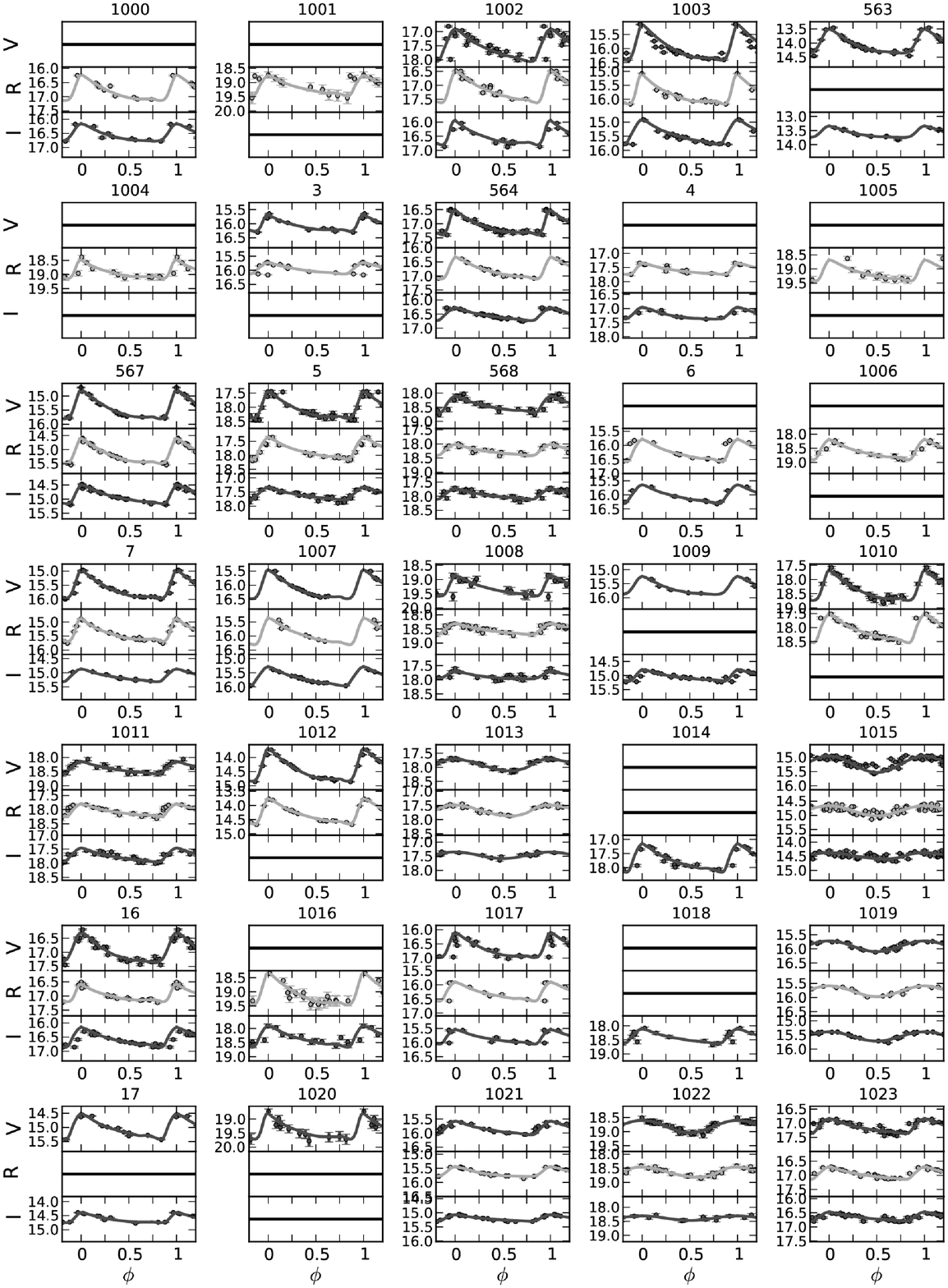} 
\contcaption{Light curves of survey \rrab~and \rrc~stars.} \label{f:rrl_lcs}
\end{center}
\end{figure*}

\begin{figure*}
\begin{center}
 \includegraphics[width=0.9\textwidth]{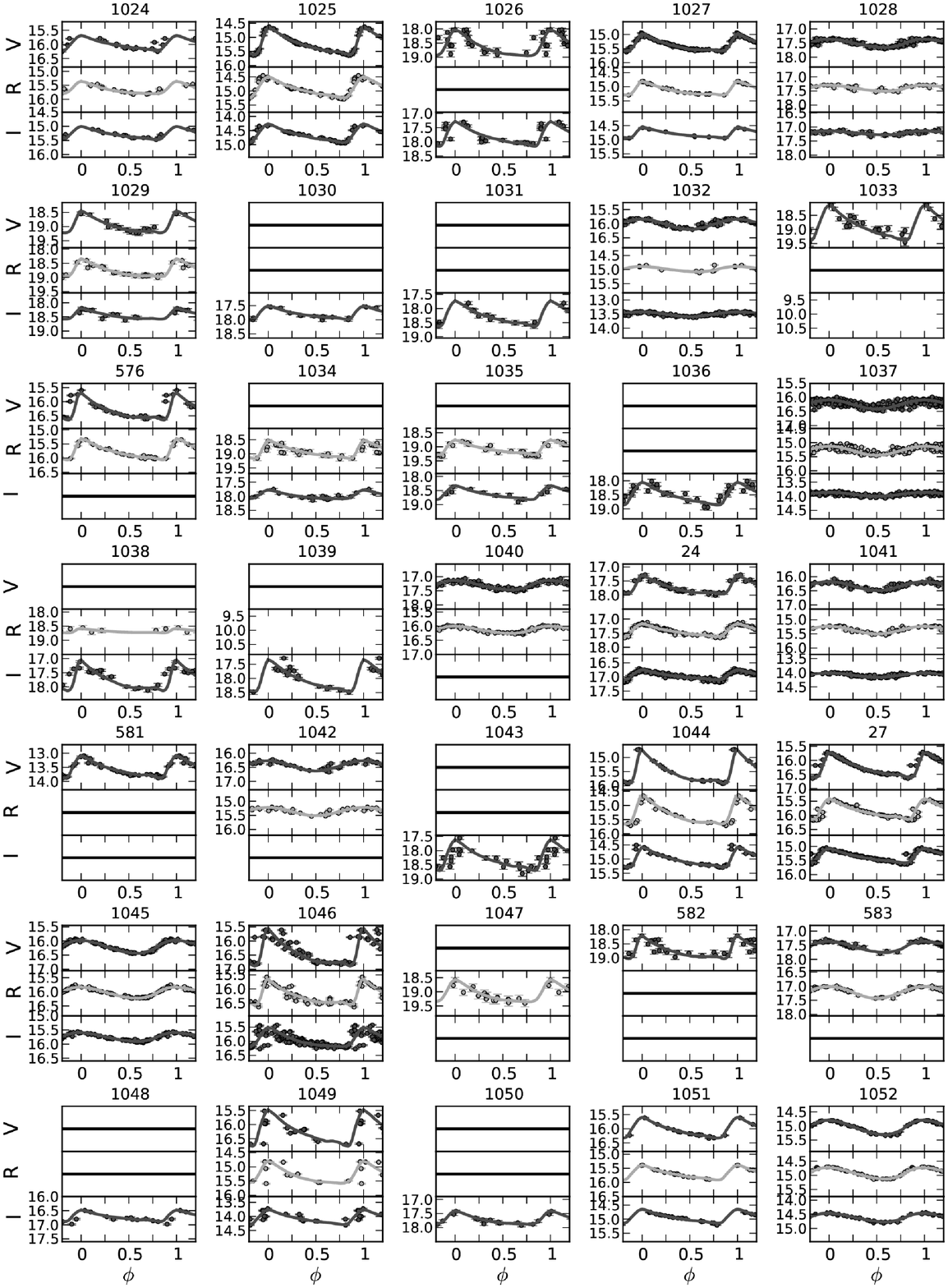} 
 \contcaption{Light curves of survey \rrab~and \rrc~stars.}
\end{center}
\end{figure*}

\begin{figure*}
\begin{center}
 \includegraphics[width=0.9\textwidth]{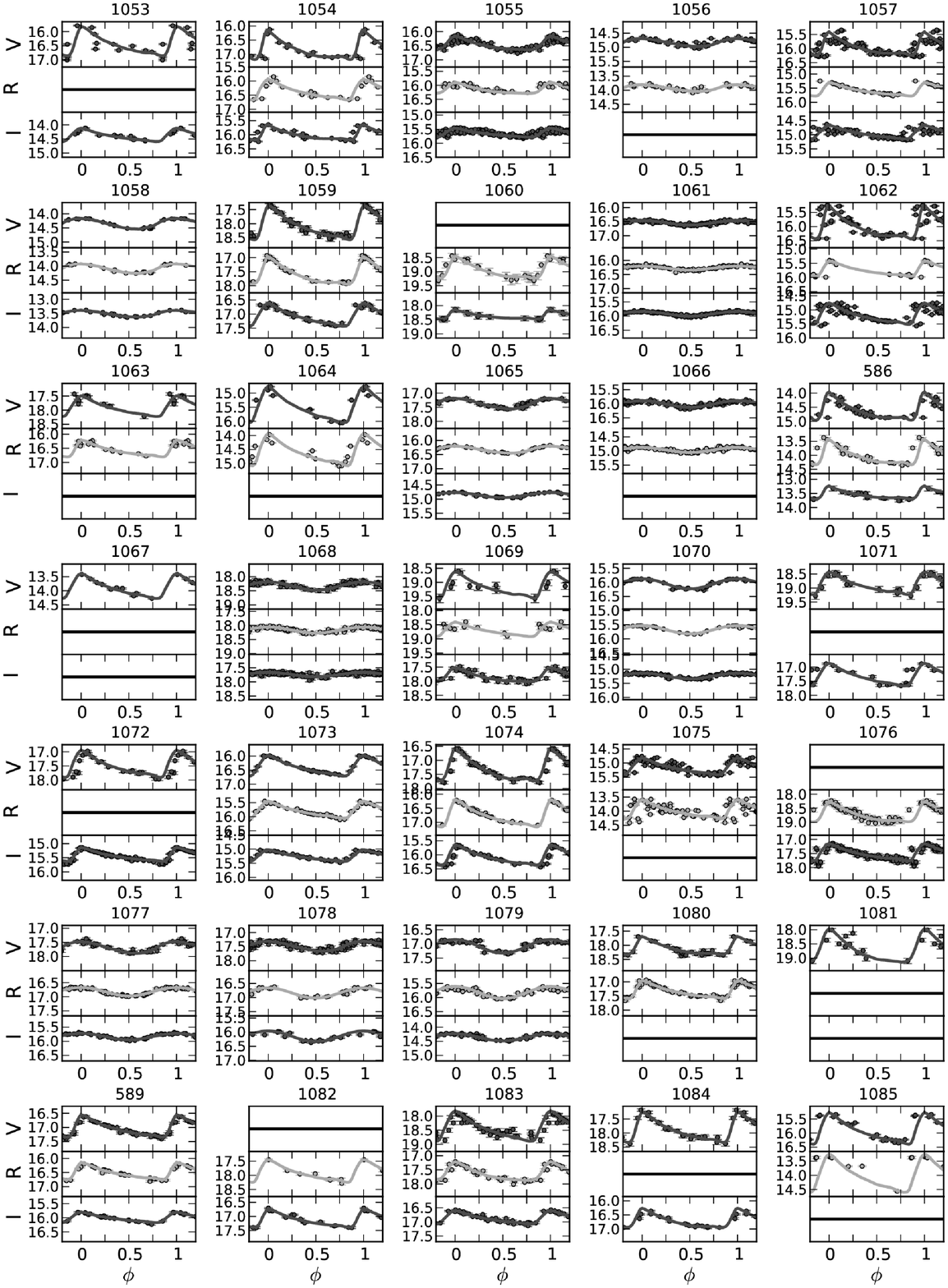} 
 \contcaption{Light curves of survey \rrab~and \rrc~stars.}
\end{center}
\end{figure*}

\begin{figure*}
\begin{center}
 \includegraphics[width=0.9\textwidth]{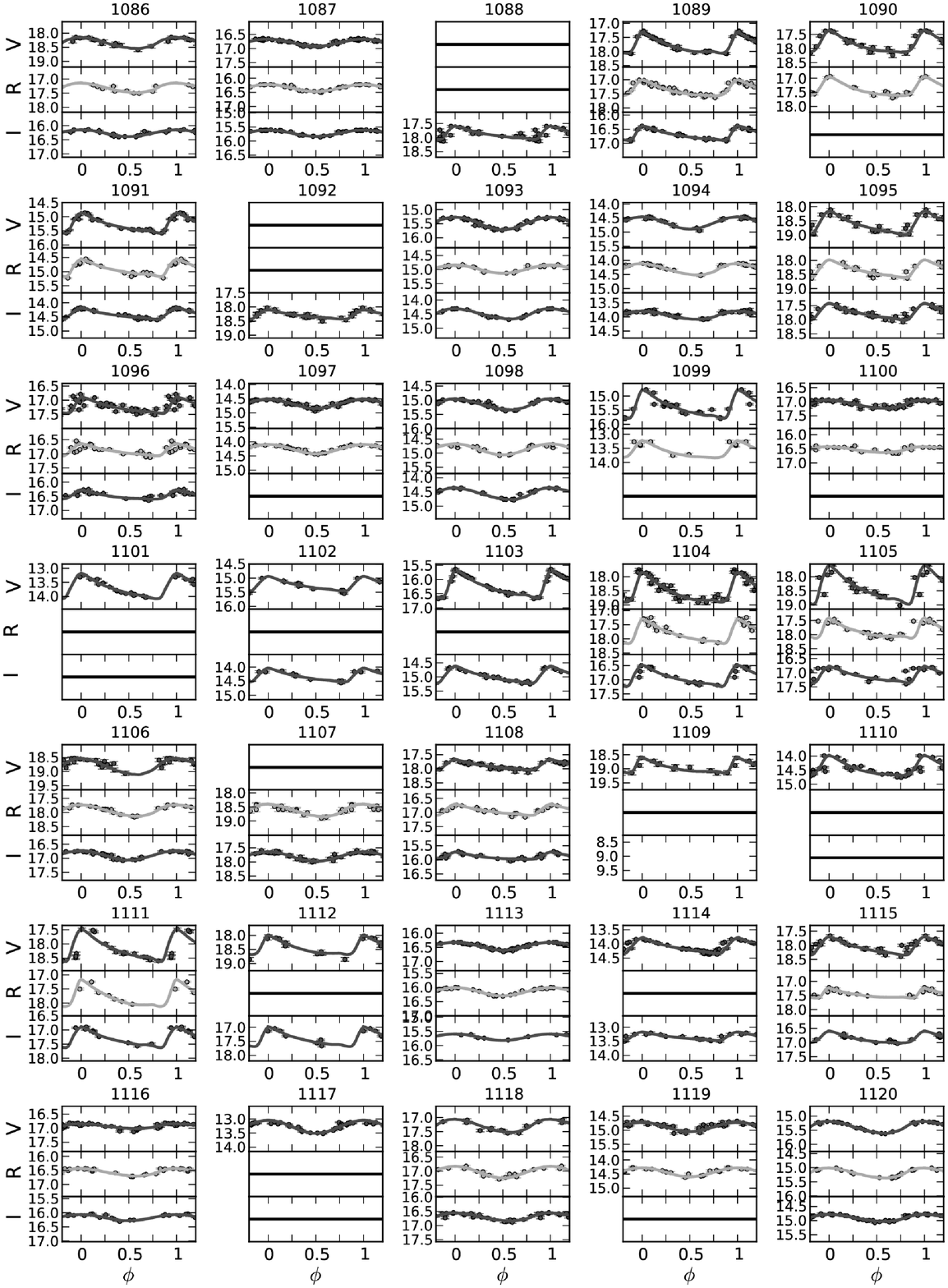} 
 \contcaption{Light curves of survey \rrab~and \rrc~stars.}
\end{center}
\end{figure*}

\begin{figure*}
\begin{center}
 \includegraphics[width=0.9\textwidth]{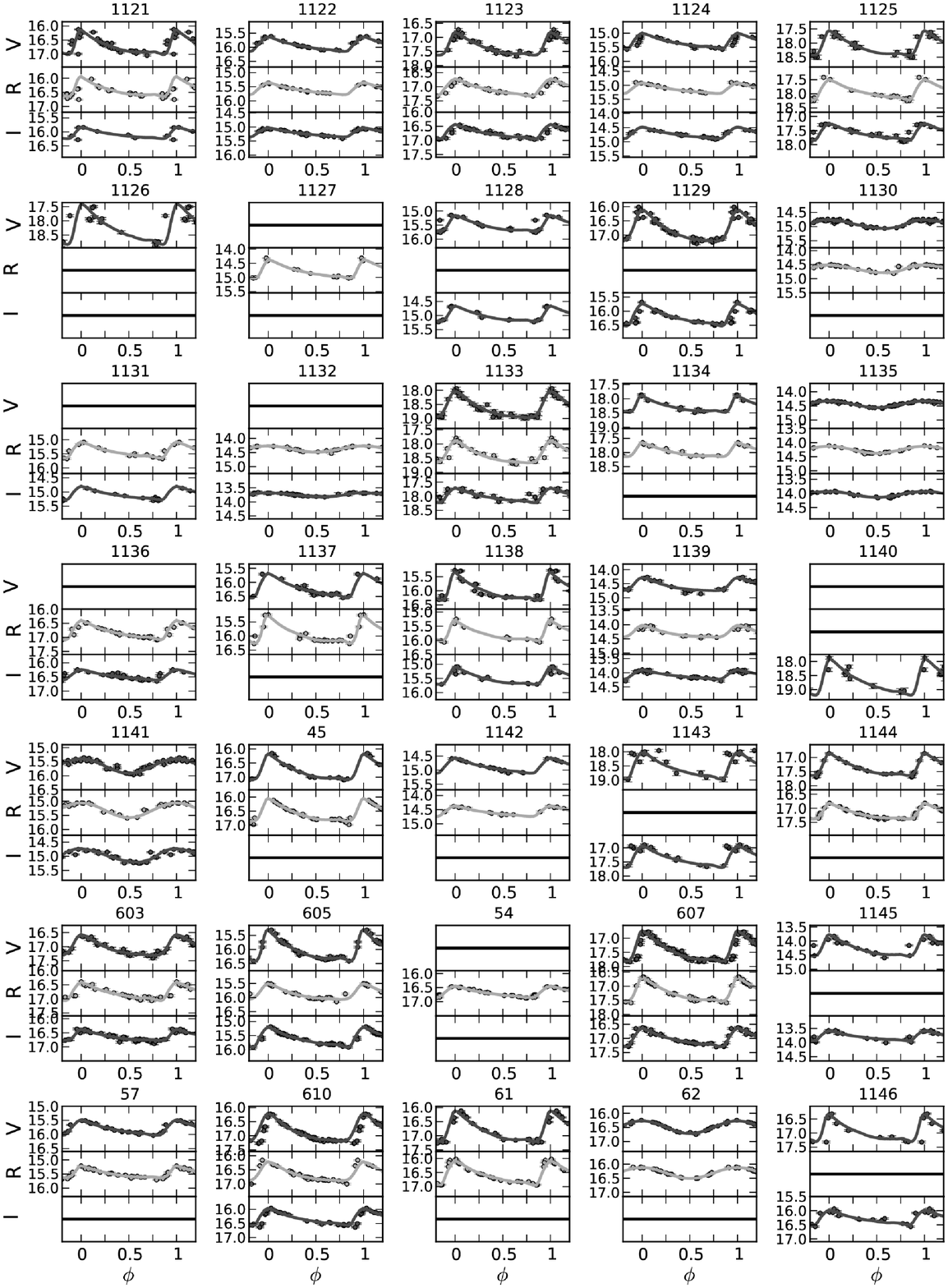} 
 \contcaption{Light curves of survey \rrab~and \rrc~stars.}
\end{center}
\end{figure*}

\begin{figure*}
\begin{center}
 \includegraphics[width=0.9\textwidth]{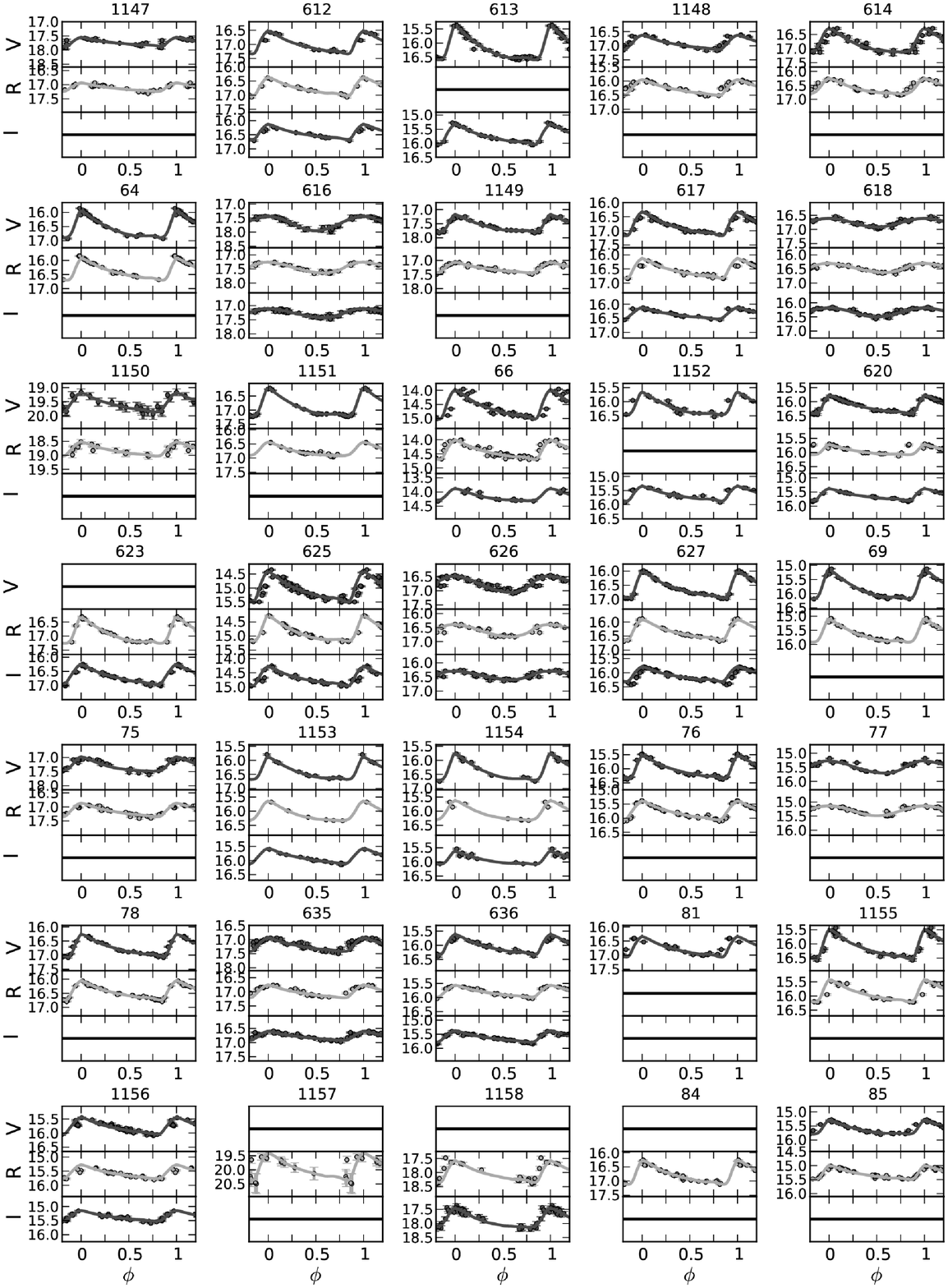} 
 \contcaption{Light curves of survey \rrab~and \rrc~stars.}
\end{center}
\end{figure*}

\clearpage

\end{document}